\documentclass[]{aastex61}
\pdfoutput=1 
\usepackage{amsmath,amstext}
\usepackage[T1]{fontenc}
\usepackage{apjfonts} 
\usepackage[figure,figure*]{hypcap}

\def\RSUN{R$_{\sun}$ }
\def\kms{$\rm km~s^{-1}$}
\def\O3{O~III]}
\def\C4{C~IV}
\def\C3{$\rm C~III$ }
\def\HE2{He~II}
\def\N5{N~V}

\def\LA{Ly$\alpha$ }

\newcommand{\hi}{H~{\sc{i}}}

\def\Si12{$\rm Si~XII$ }

\shorttitle{Coronal Magnetic Fields from Sungrazing Comets}
\shortauthors{Raymond and Giordano}

\begin{document}

\title{Probing Coronal Magnetic Fields with Sungrazing Comets: \hi\ Ly$\alpha$ from Pickup Ions}

\author{J.C. Raymond}
\affil{Harvard-Smithsonian Center for Astrophysics, 60 Garden St,
Cambridge, MA  02138, USA}

\author{S. Giordano}
\affil{INAF-Osservatorio Astrofisico di Torino, via Osservatorio 20, I-10025, Pino Torinese, Italy}

\begin{abstract}
Observations of sungrazing comets can be used to probe the solar corona, to study the composition of the comets, and to investigate the plasma processes that govern the interaction between the coronal plasma and cometary gas.   UVCS observations of the intensities and line profiles of \hi\ Ly$\alpha$ trace the density, temperature and outflow speed of the corona.  Analysis of \hi\ Ly$\alpha$ observations of comet C/2002 S2 showed a surprising split in the comet's Ly$\alpha$ tail and an asymmetry of red-shifted and blue-shifted emission across the tail axis.  It was suggested that the velocity structure might result from a population of neutrals produced by charge transfer between pickup ions and cometary neutrals.  Here we present numerical simulations of the \hi\ Ly$\alpha$ intensity and velocity centroid for sungrazing comets under the assumptions that the magnetic field and solar wind are radial.  The models qualitatively reproduce the observations of Comet C/2002 S2 and potentially explain the split tail morphology that was seen in C/2002 S2 and also C/2001 C2.  They also match the observed red- and blue-shifts, though the solar wind velocity needed to explain the blue-shift implies strong Doppler dimming and requires a higher outgassing rate to match the light curve.  However, the models do not match the observations in detail, and we discuss the remaining discrepancies and the uncertainties in the model.  We briefly discuss the implications for other UVCS comet observations and sungrazing comet observations with the Metis coronagraph. 
\end{abstract}


\section{Introduction}

The solar system contains a large number of sungrazing comets.  Their perihelia lie just above the surface of the Sun, so very few survive perihelion passage.  Most belong to the Kreutz family, and those follow the same orbital path.  More than 3000 Kreutz sungrazers have been discovered, mostly by the Large Angle and Spectrometric Coronagraph (LASCO) on the SOHO satellite.  
Marsden (2005) reviews the history and orbital evolution of sungrazers, and a recent comprehensive review is given by Jones et al.(2017).

Sungrazing comets are interesting in their own right, but they are also valuable probes of the corona and inner solar wind.  Remote sensing observations of the corona usually provide intensities integrated along the entire line of sight, from which one derives the average properties of the corona.  Observations of a sungrazing comet, on the other hand, can provide plasma properties at points along the comet trajectory without the line of sight (LOS) averaging, acting as a probe of the corona.  Observations of two large sungrazers, C/2011 N3 and C/2011 W3 (Lovejoy), by the AIA instrument on SDO, have been used to infer coronal densities and magnetic field directions (Schrijver et al.(2012), Bryans \& Pesnell (2012), McCauley et al.(2013), Downs et al.(2013), Raymond et al.(2014)).

The intensities and profiles of the Ly$\alpha$ tails of 5 sungrazers observed by the Ultraviolet Coronagraph Spectrometer (UVCS) instrument on SOHO (Kohl et al.(1997), Kohl et al.(2006)) have been interpreted in terms of the densities, temperatures and outflow speeds of the wind (Raymond et al.1998, Uzzo et al. 2001, Bemporad et al. 2005, Ciaravella et al. 2010, Giordano et al.2015). However, Comet C/2002 S2 presented a surprising anomaly.  Figures~1 and 2 show the intensity and velocity centroid images of that comet reconstructed from the UVCS spectra (Giordano et al. 2015).  The split tail observed at 8 \RSUN and the asymmetric pattern of red-shift to the south and blue-shift to the north could not be understood in the context of the models then available. The present paper explores that anomaly.

When water sublimates from the surface of the comet, UV light photodissociates the molecules, producing hydrogen atoms that form a cloud that moves with the comet and expands at about 10~\kms.  We will call these first generation neutrals.  They can scatter Ly$\alpha$ photons from the solar disk, but because they move at the speed of the comet, the absorption profile is shifted away from the disk emission profile, and the scattered intensity is reduced by Doppler Dimming (Swings effect).

The 1$^{st}$ generation neutrals are subject to collisional ionization, photoionization and charge transfer with coronal protons, with charge transfer generally dominating.  The charge transfer produces a population of neutral H atoms with approximately the bulk velocity and the thermal velocity distribution of the coronal protons, and we call these second generation neutrals. These neutrals also scatter Ly$\alpha$ photons from the disk, and their larger line width and generally smaller flow speed reduce the effects of Doppler dimming enough that this component dominates inside 10~\RSUN.  The centroid shift of the Ly$\alpha$ profile is the LOS component of the coronal (i.e., solar wind) velocity, and the profile width and the expansion rate of the cloud of 2$^{nd}$ generation neutrals are governed by the thermal width of the coronal protons.  The intensity of the 2$^{nd}$ generation emission drops off as the neutrals are ionized, so the decay time gives the local electron density.  If the local solar wind speed is comparable to the comet speed, an angular offset between the comet trajectory and the axis of the cloud of 2$^{nd}$ generation neutrals can be used to infer the $V_{wind}$ (Bemporad et al.(2015)). 

Ionization of the 1$^{st}$ generation neutrals (but not the 2$^{nd}$) produces pickup ions (PUIs).  The 1$^{st}$ generation neutrals move with the comet, and when they become ionized the velocity component perpendicular to the magnetic field becomes gyro motion around the field line, while the parallel component is conserved.  The gyro motion makes a ring beam in velocity space, which is unstable and rapidly evolves to a bispherical shell in velocity space (Williams \& Zank(1994), Isenberg \& Lee(1996)).  On a somewhat longer time scale, the shell evolves into a Maxwellian, as seen in the oxygen ions in Comet C/2011 W3 (Lovejoy) by Raymond et al.(2014). Pickup ion distributions have been measured in association with comets (Coates \& Jones(2009)) and in the ambient solar wind (Moebius et al.(1985), Gloeckler et al.(1993)). 

Charge transfer between pickup ions and either 1$^{st}$ or 2$^{nd}$ generation neutrals produces a population of neutrals with approximately the same bulk velocity and random velocity distribution as the PUIs.  That is to say that they move along the magnetic field at a speed $V_{\|}=V_{com} cos \theta$ and have a random velocity $V_{\bot}=V_{com} sin \theta$, where $V_{com}$ is the comet speed and $\theta$ is the angle between the comet velocity and the magnetic field.  
We call these third generation neutrals.

Figure~3 shows a schematic view of the 3 populations.  The 1$^{st}$ generation can be seen as a spherical cloud around the comet, the 2$^{nd}$ as a broad conical structure whose axis lies between the comet velocity vector and the magnetic field, and the 3$^{rd}$ generation as a narrower structure offset to the opposite side of the comet's path.  The diagram shows an image in the rest frame of the comet.  In the observer's frame, the 3$^{rd}$ generation neutrals move at an angle to the positive X direction, as is seen for the striations in Comet Lovejoy (Raymond et al (2014)).

The 3$^{rd}$ generation neutrals offer an explanation for the anomalies seen in the UVCS observations of comet C/2002 S2 (Giordano et al 2015).  
Giordano et al (2015) suggested that 3$^{rd}$ generation neutrals could account for both the split tail and the asymmetric line centroids, assuming that the blue-shift is the LOS component of the solar wind and the red-shift is the LOS component of the parallel velocity component of the PUIs.  However, they provided no quantitative models.

This paper presents a first attempt at a quantitative analysis of the effects of 3$^{rd}$ generation neutrals on the intensity and centroid of the Ly$\alpha$ line.  We present the model calculations in Section 2, results in Section 3 and comparisons with the observations of C/2002 S2 in Section 4, along with some discussion of the approximations and applicable parameter ranges of the models.  Section 5 provides a summary and a description of how 3$^{rd}$ generation neutrals both complicate and enhance the analysis of UVCS spectra and Metis Ly$\alpha$ images of sungrazing comets.

\begin{figure*}
	\begin{tabular}{cc}
		\resizebox{90mm}{!}{\includegraphics{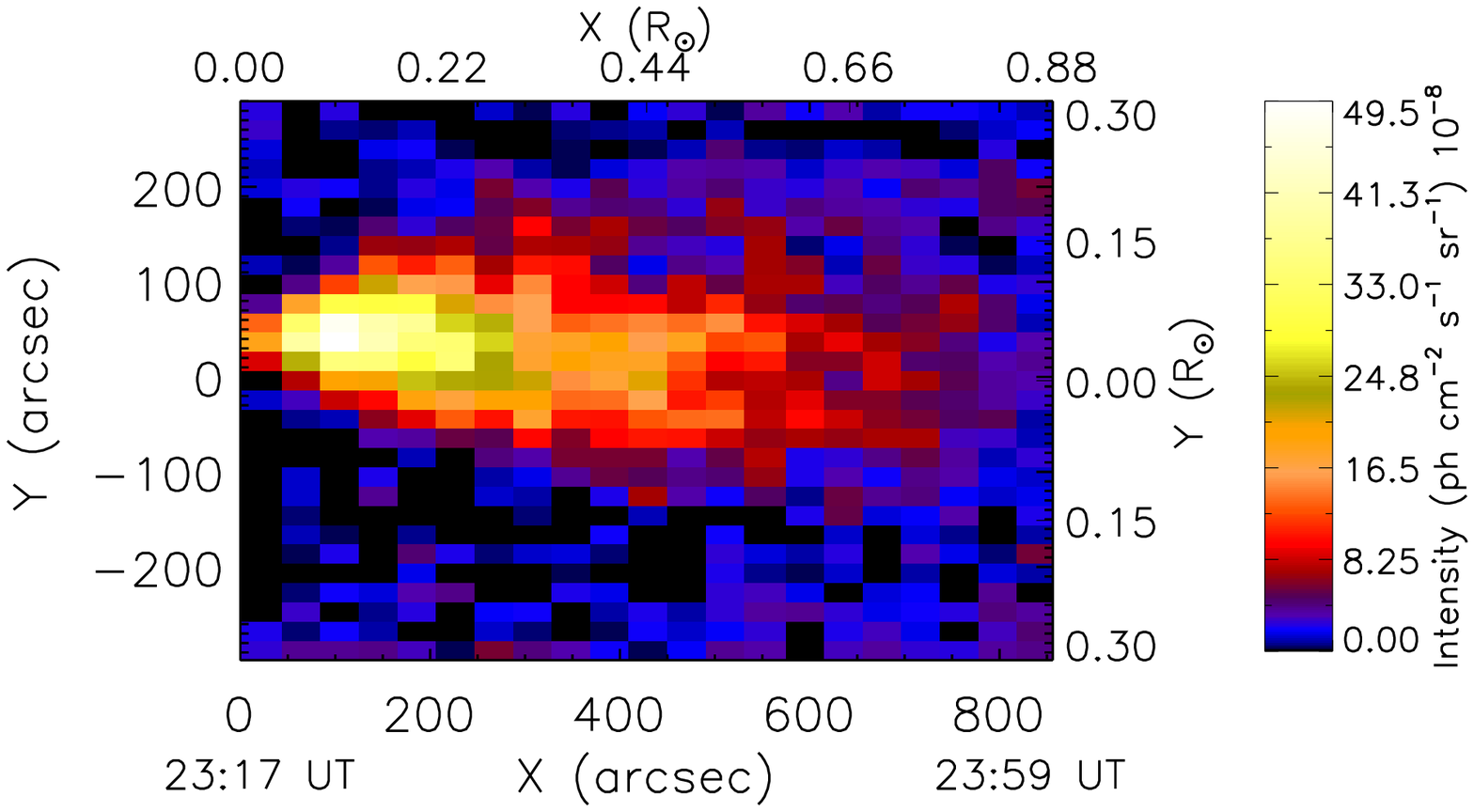}}  &	\resizebox{90mm}{!}{\includegraphics{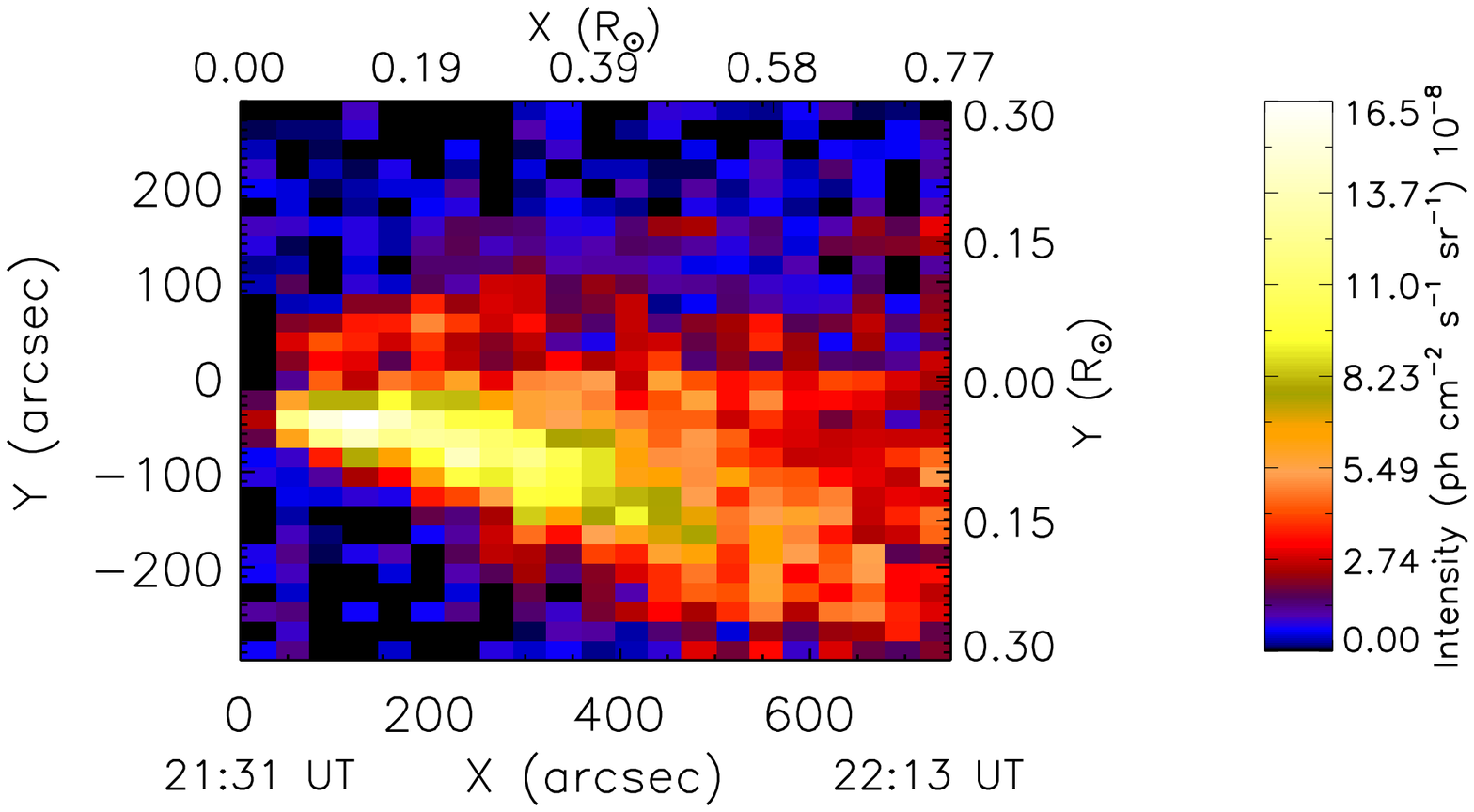}}   \\
	\end{tabular}
	\caption{Reconstructed \hi\, Ly$\alpha$ intensity images of Comet C/2002 S2, observed on 18 September, 2002, 
	as it crossed the UVCS slit at actual distances 6.00 (left panel) and 8.00~\RSUN (right panel) from Sun center (projected heliocentric distances 
	are 4.55 and 6.84~\RSUN, respectively (Giordano et al (2015)).  Initial and last UT times of the comet observation at each height are reported 
	at the left and right image corners.
	\label{giordano7}}
\end{figure*}

\section{MODELS}

We treat the comet as a test particle, which for our purposes means that the mass injected by the comet into the corona does not strongly perturb the corona.  If the outgassing rate is high, the comet can drive a bow shock through the coronal plasma (Gombosi et al. 1996).  Comet C/2011 Lovejoy with a mass loss rate of $10^{10}$ g/s near perihelion perturbed the magnetic structure of the corona as indicated by motion of the striations seen in AIA 171 \AA\/ images after the comet's passage (Raymond et al. 2014) and by dramatic changes in the velocity centroids and shifts of the \LA\/ line (Raymond et al. 2018).  More details are given in Jones et al. (2017).  Here we will consider smaller comets with correspondingly smaller outgassing rates.

Our model considers a comet moving through a uniform corona.  It ejects neutral material at a rate \.{N} hydrogen atoms per second and we follow the atoms as they scatter Ly$\alpha$ (including the effects of line scattering opacity for the 1$^{st}$ generation neutrals and Doppler dimming for all the components) and undergo ionization and charge transfer reactions.  The atoms interact with a corona whose density, temperature and outflow speed are specified.  The coronal outflow is taken to be parallel to the magnetic field, and both are expected to be approximately radial between the streamer cusps at around 4~\RSUN and the Alfv\'{e}n surface at around 10 to 20~\RSUN where the field bends into a Parker spiral.

In the model the comet positions, R in solar radii, are derived from the observed positions, and the model parameters such as the comet speed, $V_{com}$, and the Earth-Sun-Comet angle, also called the phase angle, $\alpha$, are known from the comet orbit.  Moreover, the angle between the comet trajectory and the magnetic field, $\theta$, is known from the comet orbit if the field is radial.  For comparison with observations, $\alpha$ is also important, as it determines the projection of the 3D model onto the plane of the sky and the LOS components of the velocities.  The free model parameters are the cometary outgassing rate, \.{N},  the coronal proton density, $n_p$, temperature, $T_{cor}$, and wind speed, $V_{wind}$.
The magnetic field strength is not an important parameter in our models because the structure is not perturbed by the comet.
However, the strength of B does affect the loss of energy to Alfv\'{e}n waves as the PUIs scatter from a ring beam to bispherical shell distribution (Williams \& Zank 1994, Raymond et al. 2010), but we will ignore that complication here.

\begin{table} [h]
\caption {Model parameters}
\centerline{
\begin{tabular}{|l|l|l|}
\hline
Quantity	& \multicolumn{2}{|l|}{~~~~~~~Value or Range}      \\
\hline
$R$			& 8.00~\RSUN        & 6.00~\RSUN            \\
$V_{com}$ 	& 218~\kms	        & 252~\kms              \\
$\alpha$ 	& 31.40$^{\circ}$   & 38.88$^{\circ}$       \\
$\theta$ 	& 22.14$^{\circ}$   & 25.86$^{\circ}$       \\
\hline
$\dot{N}$   & \multicolumn{2}{|l|}{6$\times 10^{27}$ to 3$\times 10^{29}$ H $\rm s^{-1}$}  \\
$n_p$       & \multicolumn{2}{|l|}{2.5$\times 10^{3}$ to 6.0$\times 10^{4}$  $\rm cm^{-3}$}  \\
$V_{wind}$      & \multicolumn{2}{|l|}{50 to 400 \kms} \\
$T_{cor}$         & \multicolumn{2}{|l|}{6.0 and 6.2 log T}    \\
$V_{1}$         & \multicolumn{2}{|l|}{10 \kms}  \\
\hline
\multicolumn{3}{|l|}{$R$: Comet heliocentric distance}              \\
\multicolumn{3}{|l|}{$V_{com}$: Comet speed}                        \\
\multicolumn{3}{|l|}{$\alpha$: Phase angle}                         \\ 
\multicolumn{3}{|l|}{$\theta$: Comet trajectory and the magnetic field angle}      \\
\multicolumn{3}{|l|}{$\dot{N}$: Cometary outgassing rate}                \\ 
\multicolumn{3}{|l|}{$n_p$: Coronal proton density}             \\ 
\multicolumn{3}{|l|}{$V_{wind}$: Solar wind speed}                  \\
\multicolumn{3}{|l|}{$T_{cor}$: Coronal proton temperature}                 \\
\multicolumn{3}{|l|}{$V_{1}$: 1$^{st}$ generation outflow speed}    \\
\hline
\end{tabular}
}
\end{table}

\subsection{Atomic Rates}

To compute the rate of Ly$\alpha$ scattering we precompute a lookup table of the number of photons scattered per second per H atom as a function of the flow speed in the radial direction and thermal width, then scale by the dilution factor of the solar radiation at the heliocentric height of the comet.  For the integration over frequency and the illuminating disk, we use a calculation by S. Cranmer (private communication).  We use a Ly$\alpha$ disk intensity appropriate for solar minimum and increase it by a factor of about 1.8 for solar maximum conditions.  For 1$^{st}$ generation neutrals we include the opacity in the Ly$\alpha$ line using the outflow speed of the 1$^{st}$ generation neutrals, $V_1$, for the Doppler width, and we reduce the scattering rate accordingly.  $V_1$ is determined by the energy liberated during photodissociation of $\rm H_2O$ and OH, and ranges from about 8 to 24~\kms\, for the H atoms (Shimizu (1991).  For the opacity calculation, we assume that B is radial.

The ionization rate is given by

\begin{equation}
q_{ion} = n_p q_{ex} ~+~ n_e q_i ~+~ q_{phot}
\end{equation}

\noindent
where $q_{ex}$ is the charge transfer rate coefficient, $q_i$ is the ionization rate coefficient for electron impact, $q_{phot}$ is the photoionization rate, and $n_p$ and $n_e$ are the coronal proton and electron densities (we use the approximation $n_p=0.83n_e$ valid for solar coronal plasma).  
The charge transfer rate is computed with the cross sections of Schultz et al. (2008) with the average velocity.  The effective speed for charge transfer between 1$^{st}$ or 2$^{nd}$ generation neutrals and coronal protons or PUIs is given by the sum in quadrature of the relative bulk speed of the two components and the random (thermal) speeds of the two components. For instance, the effective speed for charge transfer between 1$^{st}$ generation neutrals and coronal protons is given by

\begin{equation}
V_{eff} = \sqrt{V_{rel}^2 + V_T^2 +V_1^2}
\end{equation}

\noindent
where $V_T$ is the thermal speed of the coronal protons (computed as the most probable speed of protons at $T_{cor}$) 
and $V_{rel}$ is the bulk speed of the coronal gas relative to the comet.  $V_1$ is the outflow speed of the 1$^{st}$ generation neutrals, which is small compared to the other terms.  We ignore charge transfer between 2$^{nd}$ generation neutrals and coronal protons, because that does not change the number of 2$^{nd}$ generation neutrals or their speed distribution.  For charge transfer between 3$^{rd}$ generation neutrals and coronal protons, the effective speed for charge transfer becomes

\begin{equation}
V_{eff} = \sqrt{V_{rel}^2 + V_T^2 + V_\bot^2 }
\end{equation},

\noindent
For the 3$^{rd}$ generation, the value of $V_{rel}$ uses the parallel velocity component of the PUIs and the solar wind speed.  

For collisional ionization, we use the ionization rate from Scholz \& Walters (1991) at an electron temperature of log~T~=~6.2.  It varies by only 3\% over the range log T=6.0-6.6.  For photoionization, we scale the rate at 6.8~\RSUN from Raymond et al. (1998) with the dilution factor as a function of heliocentric distance. This is the value appropriate for solar minimum, and we increase it by a factor of 2 for solar maximum conditions.

\subsection{Physical Picture}

The model uses a Cartesian grid, with the comet moving along the X-axis and the magnetic field in the XZ plane.  The models can then be rotated to give Doppler shifts and foreshortening for comparison with observations.  Neutrals are generated at the comet and followed as they scatter Ly$\alpha$ and undergo charge transfer until they become ionized or leave the simulation box. We use analytic descriptions of the distributions of 1$^{st}$ and 2$^{nd}$ generation neutrals.  PUIs are produced by ionization of 1$^{st}$ generation neutrals (or by ionization of 3$^{rd}$ generation neutrals).  3$^{rd}$ generation neutrals are produced by charge transfer between PUIs and 1$^{st}$ or 2$^{nd}$ generation neutrals at each grid cell, and assumed to move in the direction of the magnetic field at $V_\|$ and to spread out with a speed $V_\bot$.

We approximate the 1$^{st}$ generation distribution as a spherically symmetric outflow moving with the comet and expanding at a constant speed, with an exponential cutoff due to ionization.  Its density is

\begin{equation}
n_1(r) = \frac{\dot{N}}{4 \pi r^2 V_1} e^{-\frac{q_{ion}}{V_1}r}
\end{equation}

\noindent
where $V_1$ is the initial outflow speed due mostly to the 8 to 24 \kms\/ speeds of the H atoms ejected during photodissociation.  At high densities, the H atoms share some of their momentum with O atoms, so $V_1$ can be smaller.

The 2$^{nd}$ generation neutrals form a conical structure whose axis lies between the comet trajectory and the solar wind direction.  The angle between the comet trajectory and the 2$^{nd}$ generation axis is given by 

\begin{figure*}
	\begin{tabular}{cc}
		\resizebox{90mm}{!}{\includegraphics{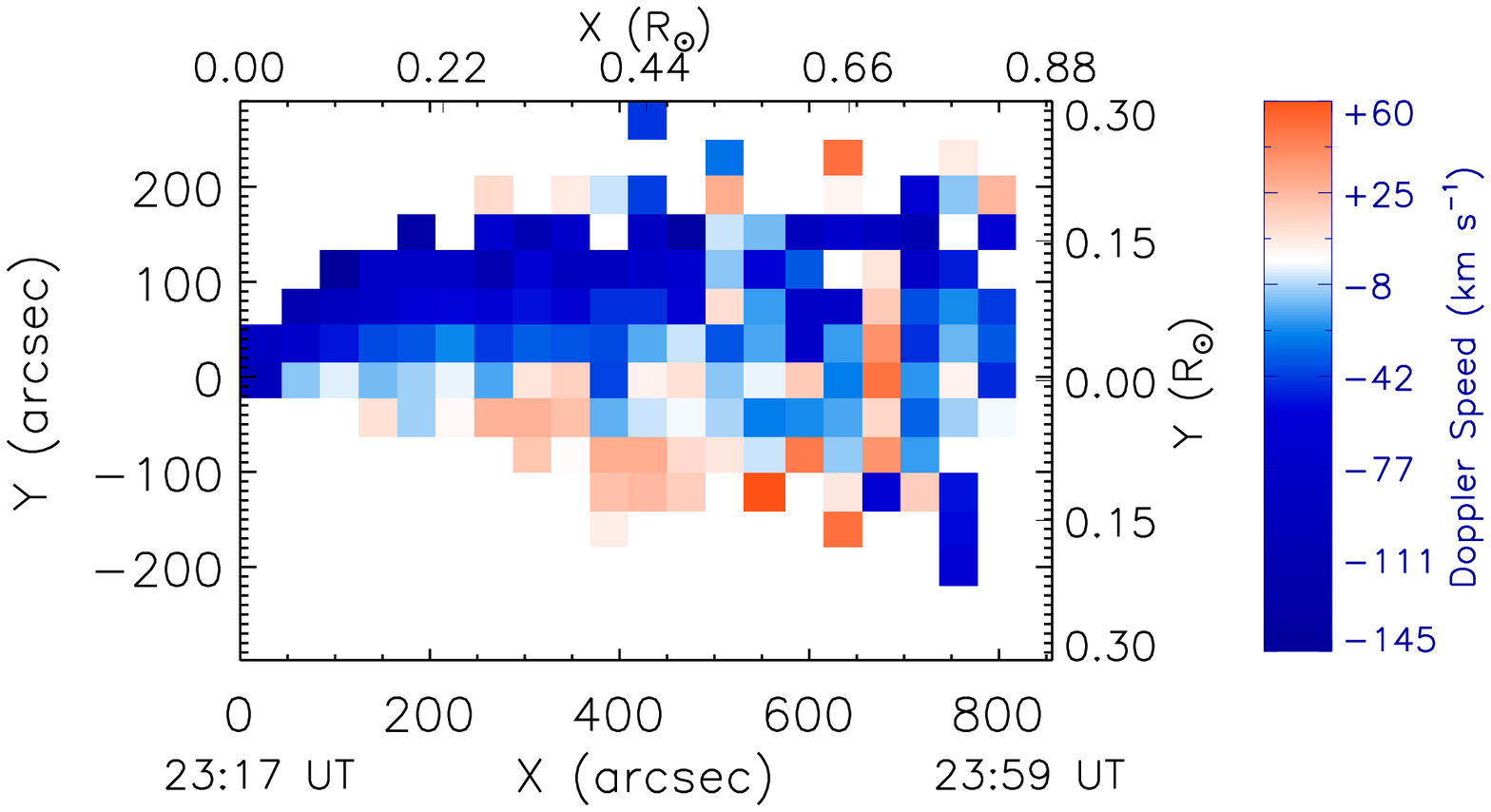}}  &	\resizebox{90mm}{!}{\includegraphics{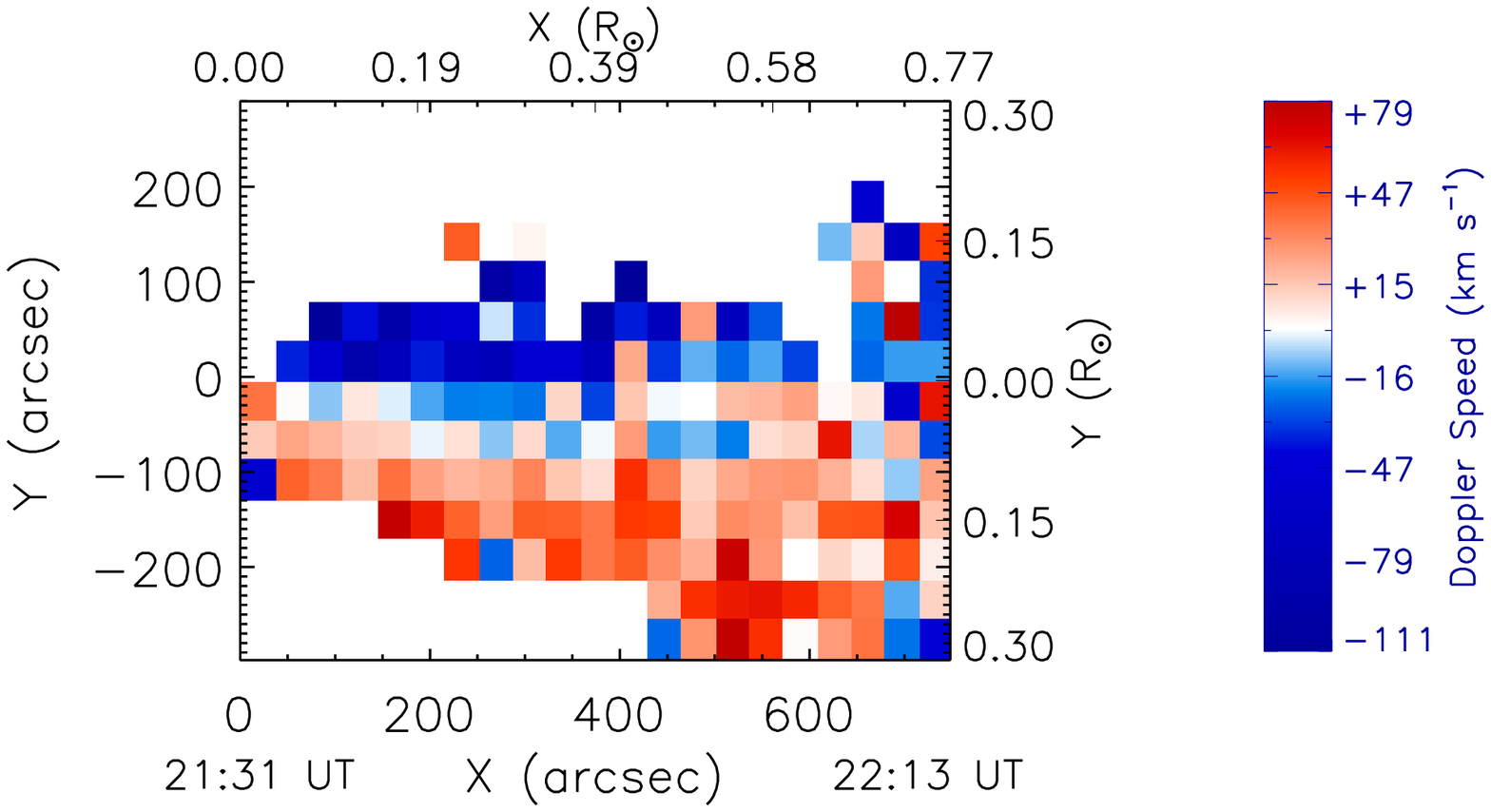}}
	\end{tabular}
	\caption{Reconstructed \hi\, Ly$\alpha$ velocity centroid images of Comet C/2002 S2, observed on 18 September, 2002, 
	as it crossed the UVCS slit at actual distances 6.00 (left panel) and 8.00~\RSUN (right panel) from Sun center (projected heliocentric distances 
	are 4.55 and 6.84~\RSUN, respectively (Giordano et al (2015)). Initial and last UT times of the comet observation at each height are reported 
	at the left and right image corners.
label{giordano9}}
\end{figure*}

\begin{equation}
sin(\beta) ~=~ V_{wind,\bot} / V_{rel}
\end{equation}

\noindent
where $V_{wind,\bot}$ is the solar wind speed component perpendicular to the comet trajectory.  We approximate the 2$^{nd}$ generation neutrals as a cone with a density falling off with distance from the axis based on a Gaussian distribution with the coronal thermal speed $V_T$.  The tail fades away on a length scale $V_{rel}/(q_{i}+q_{phot})$.

The pickup ions, and therefore the 3$^{rd}$ generation neutrals, form a structure that moves along the B field with $V_\| = V_{com} sin \theta$.  Since $V_\|$ is less than $V_{com}$ and has a component perpendicular to the comet trajectory, the PUI structure appears to trail behind the comet at an angle, displaced to the side opposite to that of the 2$^{nd}$ generation neutrals.  Because the cloud of 1$^{st}$ generation neutrals is fairly small owing to the relatively small $V_1$ and the PUIs are tightly coupled to the field lines on which they form, the main PUI structure is narrow.  The 3$^{rd}$ generation neutrals, however, can cross field lines, and they have a randomly directed velocity $V_\bot$, so the 3$^{rd}$ generation neutrals form a wider structure around the PUI field lines. 

\subsection{Numerical Approach}

The comet moves across the model grid at one cell per time step.  The 1$^{st}$ and 2$^{nd}$ generation densities are analytic functions that move with the comet.  At each time step the number of pickup ions generated at each point is computed from the density of 1$^{st}$ generation neutrals and the total (ionization plus charge transfer) ionization rate, and the pickup ions are advected along the magnetic field at $V_{\|}$ and allowed to diffuse along it at $V_{\bot} /\sqrt{3}$.  They are assumed to be unable to move across field lines.  The densities of PUIs and 1$^{st}$ and 2$^{nd}$ generation neutrals are used to compute the formation rate of 3$^{rd}$ generation neutrals, and they also advect along the field lines, but they are allowed to diffuse in 3 dimensions with a speed $V_{\bot}$.

Once the densities of the different components are known, the Ly$\alpha$ emissivity at each grid point is computed using the central velocity and line width to compute the scattering rate considering Doppler Dimming and distance from the Sun.  At each grid point the 1$^{st}$ component emission is reduced by $e^{-\tau}$, where $\tau$ is the optical depth between the grid point and the Sun. The 2$^{nd}$ and 3$^{rd}$ generation neutrals are Doppler shifted away from the 1$^{st}$ component, and they generally have much larger line widths, so they are much less absorbed. Since the velocity centroid of each component is known, the average velocity of the scattered photons is the weighted average of the velocities of the 3 components.  For comparison with observations, we rotate the model to the angle between the comet path and the line of sight, then integrate to produce a 2D projection.

Figure~4 shows a model for the parameters R=8.0~\RSUN, $V_{wind}$~=~200~\kms and $n_e$~=~$10^4~\rm cm^{-3}$,
with $\theta$ = 23$^\circ$. The intensity images show a very bright spherical 1$^{st}$ component, a conical 2$^{nd}$ component and a 3$^{rd}$ component that juts off to one side. 

\subsection{Approximations and Limitations}

We have computed approximate models to show the overall nature of the spatial and velocity structure of the Ly$\alpha$ emission from a comet.
We assume that the comet does not perturb the corona in a major way, such as inducing a bow shock like those measured near larger comets (Gombos et al 19i96).  One criterion is that the amount of mass produced by the comet be small compared with the amount of coronal mass it interacts with.  The former is just the outgassing rate.  The amount of coronal mass swept up per second is

\begin{equation}
\dot M ~=~ 4 \pi r_i^2 \mu_c n_{p} V_{rel}
\end{equation}

\noindent
where $r_i$ is the interaction length scale for ionization of cometary neutrals, $V_1 /q_i$, and $\mu_c$ is the mean weight of the coronal nuclei (Jones et al., 2017).  This gives a $\dot N~<~10^{35} V_{200} / n_p$ for $V_1=10$~\kms, where $V_{200}$ is the comet speed in units of 200~\kms .  This is consistent with the parameter ranges of interest, but the next level of approximation would be very challenging.  We defer it to a future investigation because it would require at least an MHD code of the level of sophistication of the ones used by Gombosi et al. (1996) and Jia et al.(2014).    

\begin{figure*}
\begin{center}
\resizebox{80mm}{!}{\includegraphics{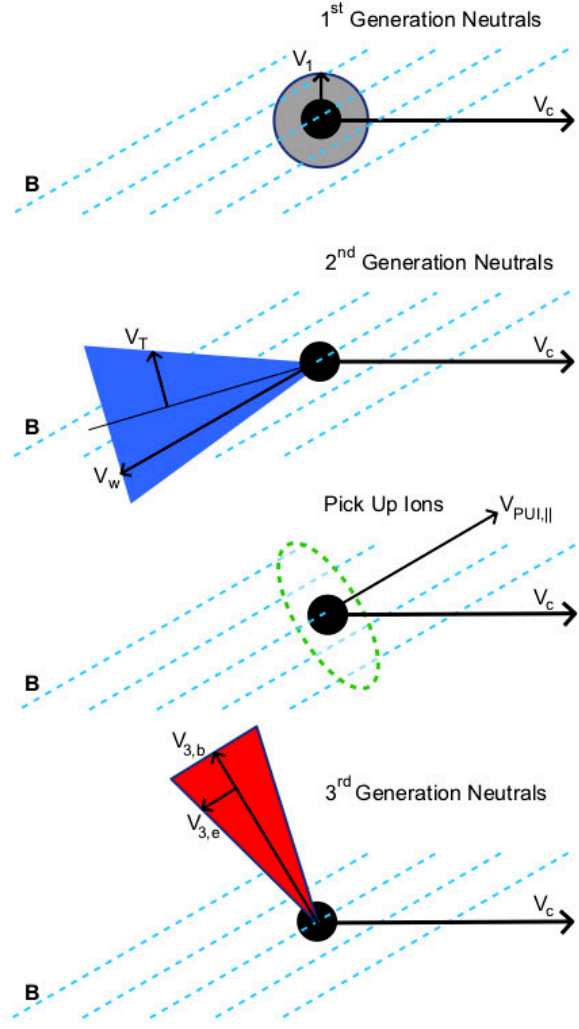}} 
\caption{Schematic diagram of the 1$^{st}$, 2$^{nd}$ and 3$^{rd}$ generation clouds. The 1$^{st}$ generation particles form a cloud, slowly expanding at $V_1$ speed, that moves with the comet.  The 2$^{nd}$ generation is a cone that moves with the wind speed in the observer's frame, or the relative speeds of the wind and the comet in the comet frame, the 2$^{nd}$ generation cone expands at proton thermal speed, $V_T$.  The Pickup ions move along the B field with the parallel component of the comet speed.  The 3$^{rd}$ generation neutrals share the pickup ion velocity, but seen in the comet frame they are moving at cos$^2$($\theta$) times the comet speed along the comet direction and sin($\theta$)cos($\theta$) in the perpendicular direction. The 3$^{rd}$ generation cloud diffuses along the B field at speed $V_{3,e}= V_{\bot}/\sqrt{3}$.
\label{schematic}}
\end{center}
\end{figure*}

Our assumption that the PUIs can be described by a shell in velocity space moving along the magnetic field at $V_\|$ is probably reasonable for perpendicular geometries, but if $\theta < 45^\circ$ streaming instabilities are likely to set in that would drive the PUI distribution to a Maxwellian and couple it to the coronal plasma.  We also assume a spherical shell in velocity space rather than the bispherical shell predicted by theory, and we neglect the loss of energy from PUIs to Alfv\'{e}n waves, which depends on the Alfv\'{e}n speed and $\theta$.

Our method of advancing time does not allow 2$^{nd}$ generation neutrals to get ahead of the comet, which can happen in $V_{com}$ and $V_{wind}$ are small compared to $V_T$.  This is not too serious a problem for most parameters, because $V_{com}$ is large close to the Sun, and $V_{wind}$ becomes larger farther away from the Sun.

We also neglect the possible production of 4$^{th}$ and higher generation neutrals by subsequent charge transfer events.  The probability of such subsequent charge transfers increases with density, and therefore with $\dot N$, so the assumption that the comet does not disturb the corona may break down before the neglect of the 4$^{th}$ and higher generations becomes a serious problem. Similarly, we neglect the loss of 1$^{st}$ and 2$^{nd}$ generation neutrals to charge transfer with PUIs, which can become important at high $\dot N$.

To address these limitations, we would probably need an MHD model that includes a neutral fluid, as in Jia et al. (2014), along with a PUI fluid.  The model from such a code would serve as the basis for a Monte Carlo calculation of the generation and transport of the neutrals and a calculation of the scattering of Ly$\alpha$.  

\section{Model Results}

Figure~4 shows the column density for the 3 neutral hydrogen populations from a typical model where the orbital path and the wind direction both lie in the plane of the sky.  This model assumed an angle between the comet's path and the wind/magnetic field direction of $\theta$ = 21.92$^{\circ}$, an outgassing rate of $10^{29}~\rm s^{-1}$, a coronal density of $10^4~\rm cm^{-3}$, a comet speed of 240~\kms\/ and wind speed of 200~\kms. The outflow speed from the comet is taken to be 10~\kms, and two proton coronal temperatures are assumed: log T$_{cor}$=6.0 and log T$_{cor}$=6.2.  
The 1$^{st}$ generation neutrals form a spherical cloud centered on the nucleus, but the emission is somewhat modified by the \LA\/ optical depth.  The 2$^{nd}$ generation atoms form a cone whose axis lies between the comet trajectory and the wind direction, and the 3$^{rd}$ generation forms a narrower tail on the other side of the comet's path.

We expect some simple approximate scalings for the total numbers of 1$^{st}$, 2$^{nd}$ and 3$^{rd}$ generation neutrals, $N_1$, $N_2$ and $N_3$.  The numbers of 1$^{st}$ and 2$^{nd}$ generation neutrals should scale as the outgassing rate times the ionization time, and therefore as $\dot{N}/n_p$.  The number of 3$^{rd}$ generation particles is proportional to the number of 1$^{st}$ generation neutrals times the number of PUIs multiplied by the size of the interaction region. The coronal density drops out because it appears in both ionization time of the 3$^{rd}$ generation particles and in the size of the interaction region.  From the numerical models we have approximately

\begin{equation}
N_1~~=~~8 \times 10^{31}  ~\dot{N}_{29} / n_4 \\
\end{equation}
\begin{equation}
N_2~~=~~6 \times 10^{31}  ~\dot{N}_{29} / n_4 \\
\end{equation}
\begin{equation}
N_3~~=~5 \times 10^{30}  ~\dot{N}_{29}^2 / V_{10}
\end{equation}

\begin{figure*}[h]
  \begin{center}
    \begin{tabular}{ccc}
      \resizebox{60mm}{!}{\includegraphics{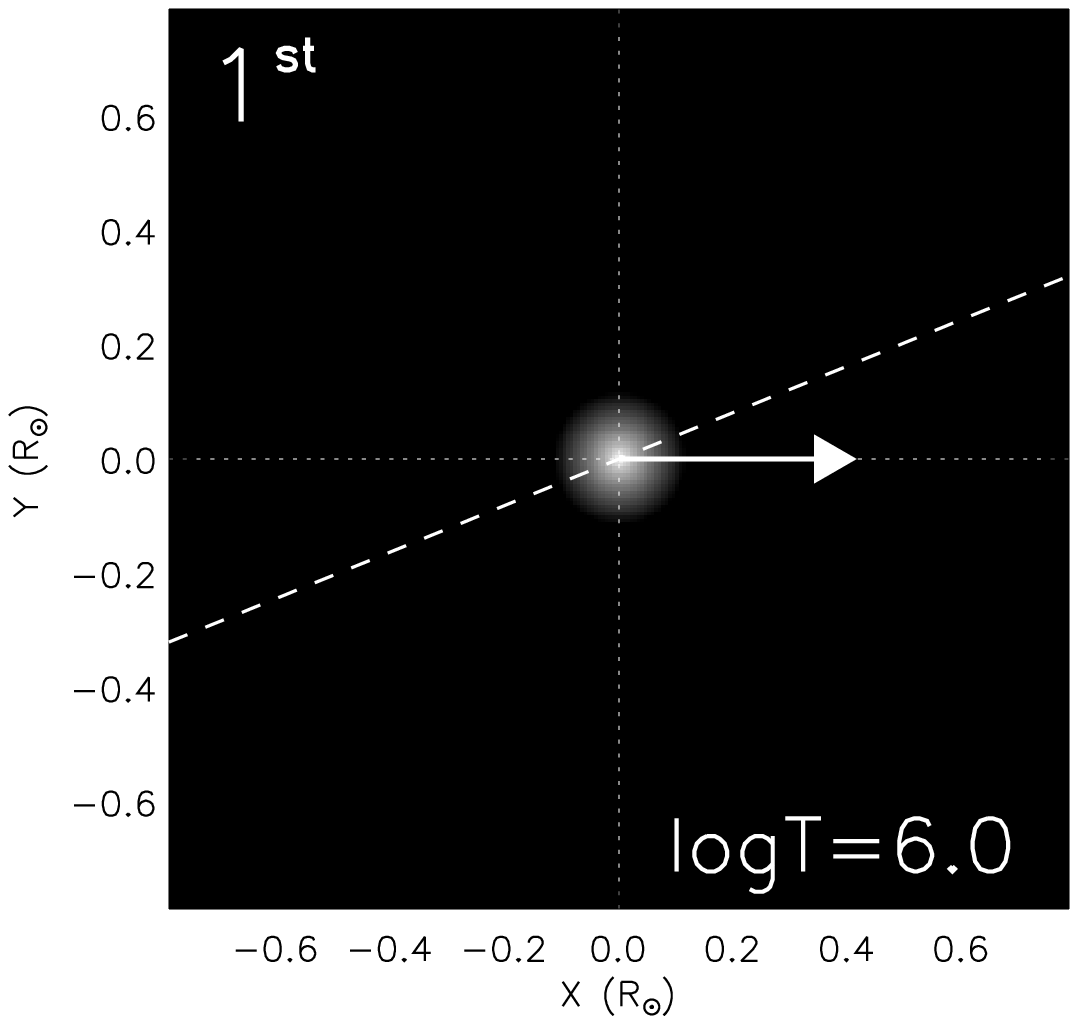}}  &
      \resizebox{60mm}{!}{\includegraphics{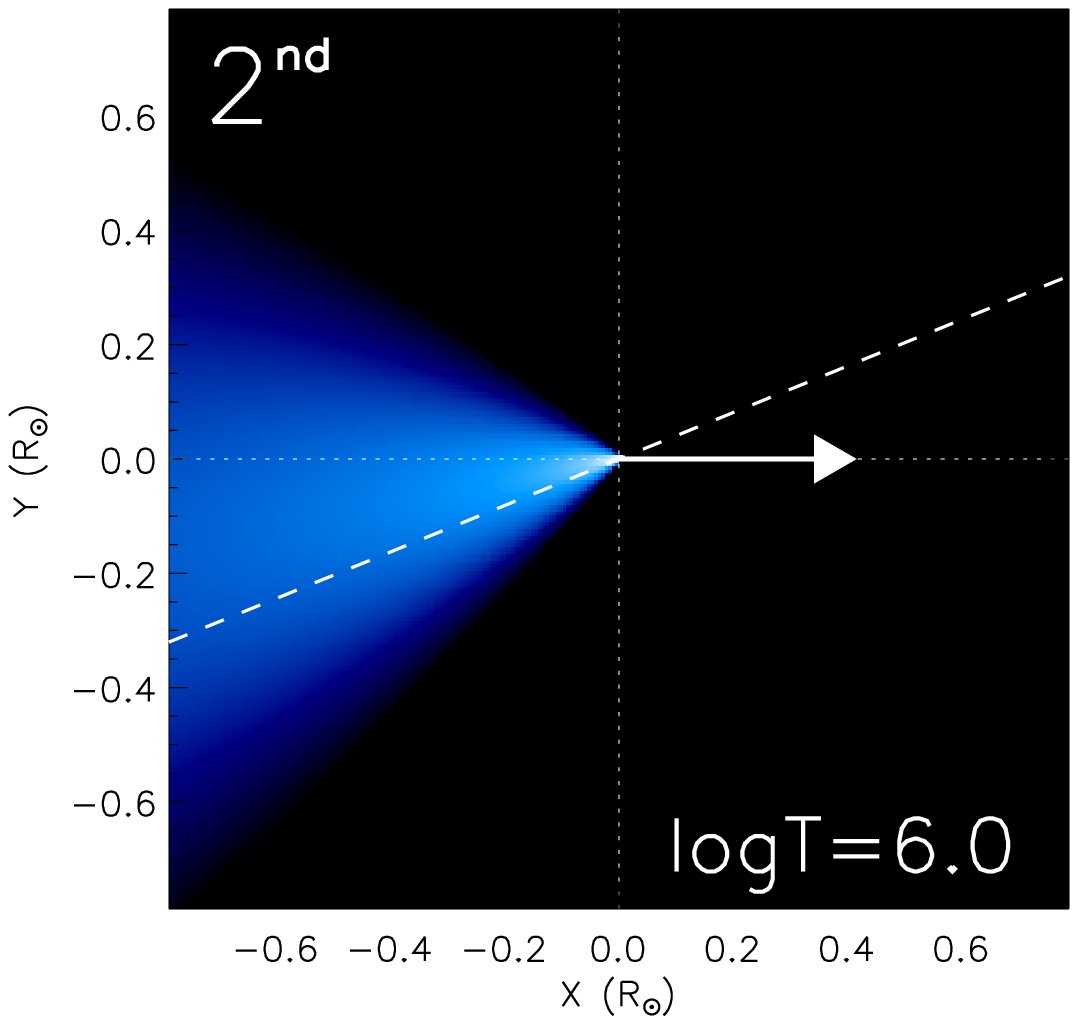}}  &
      \resizebox{60mm}{!}{\includegraphics{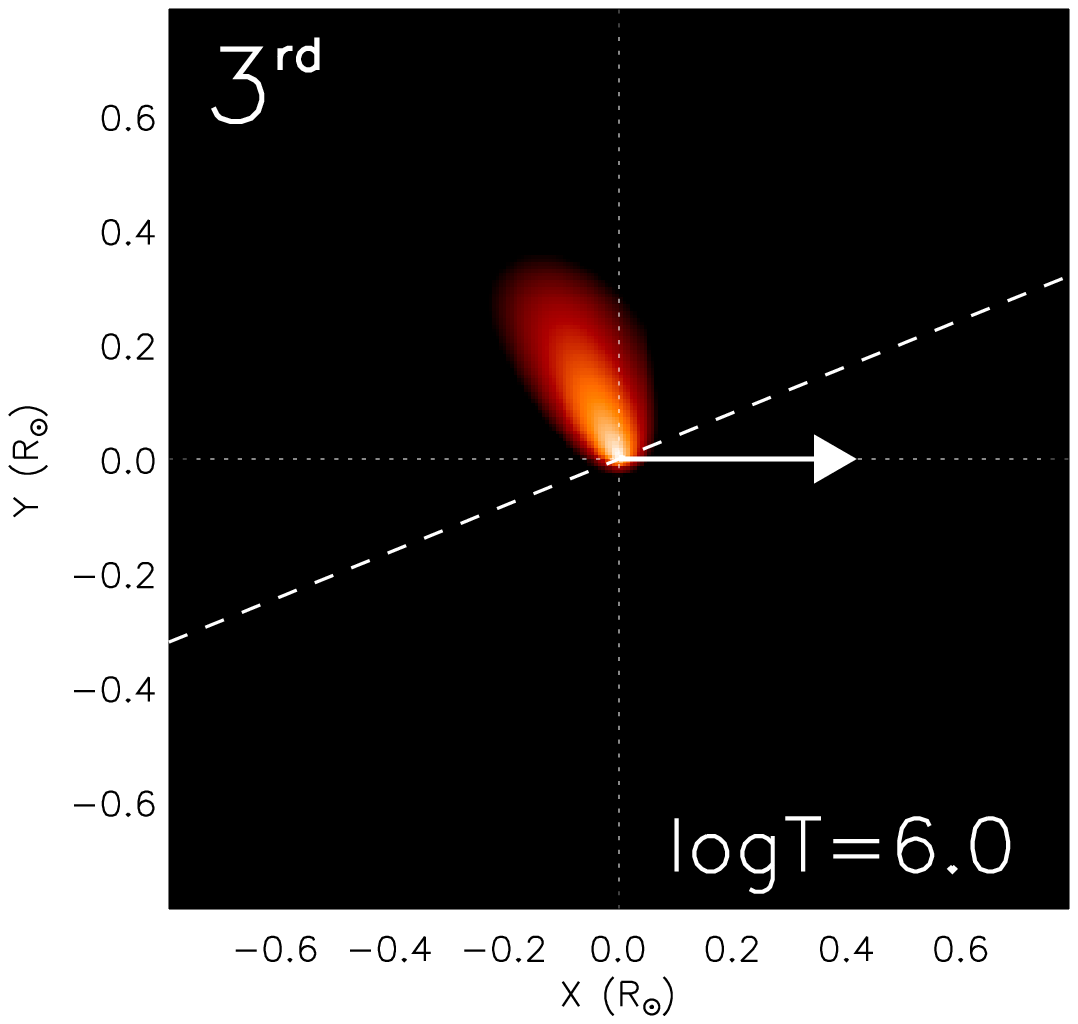}}  \\
      \resizebox{60mm}{!}{\includegraphics{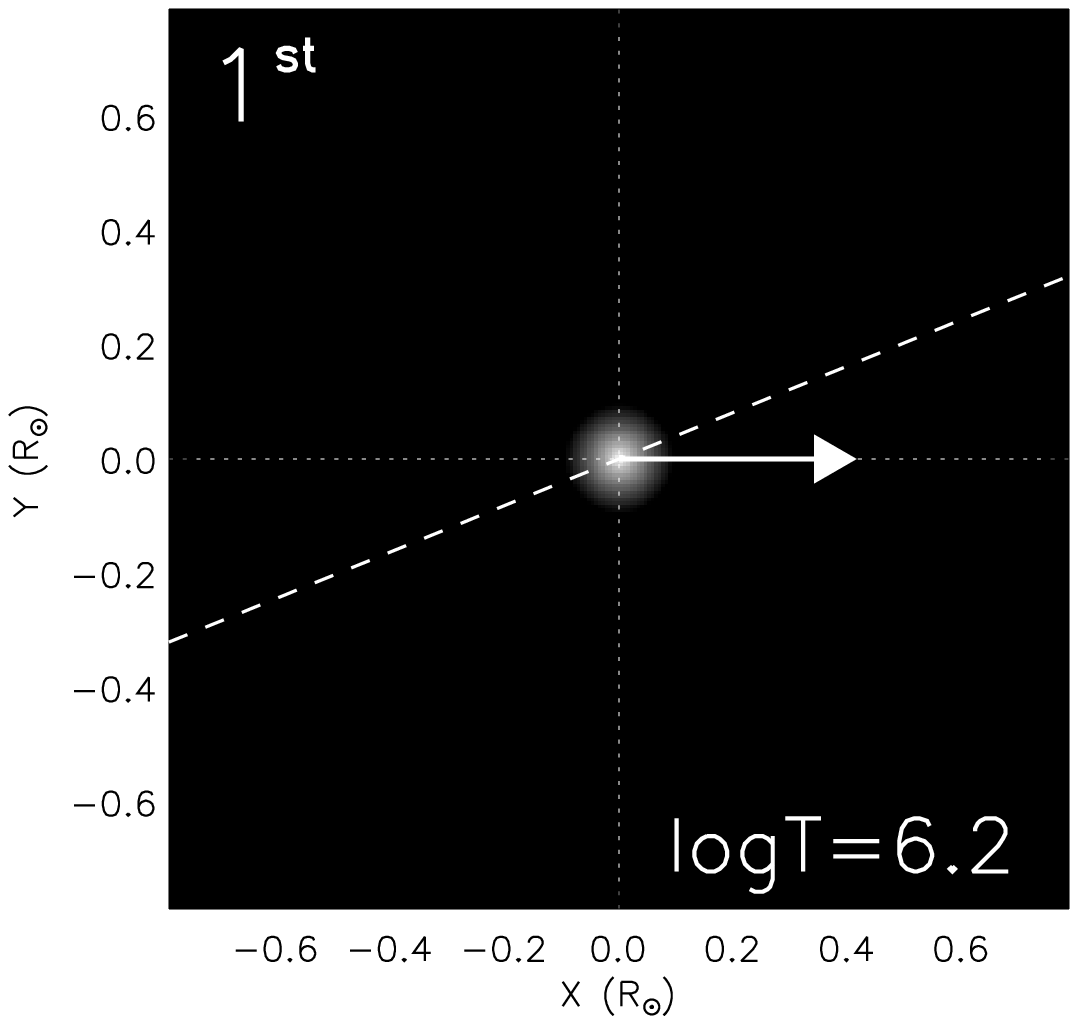}}  &
      \resizebox{60mm}{!}{\includegraphics{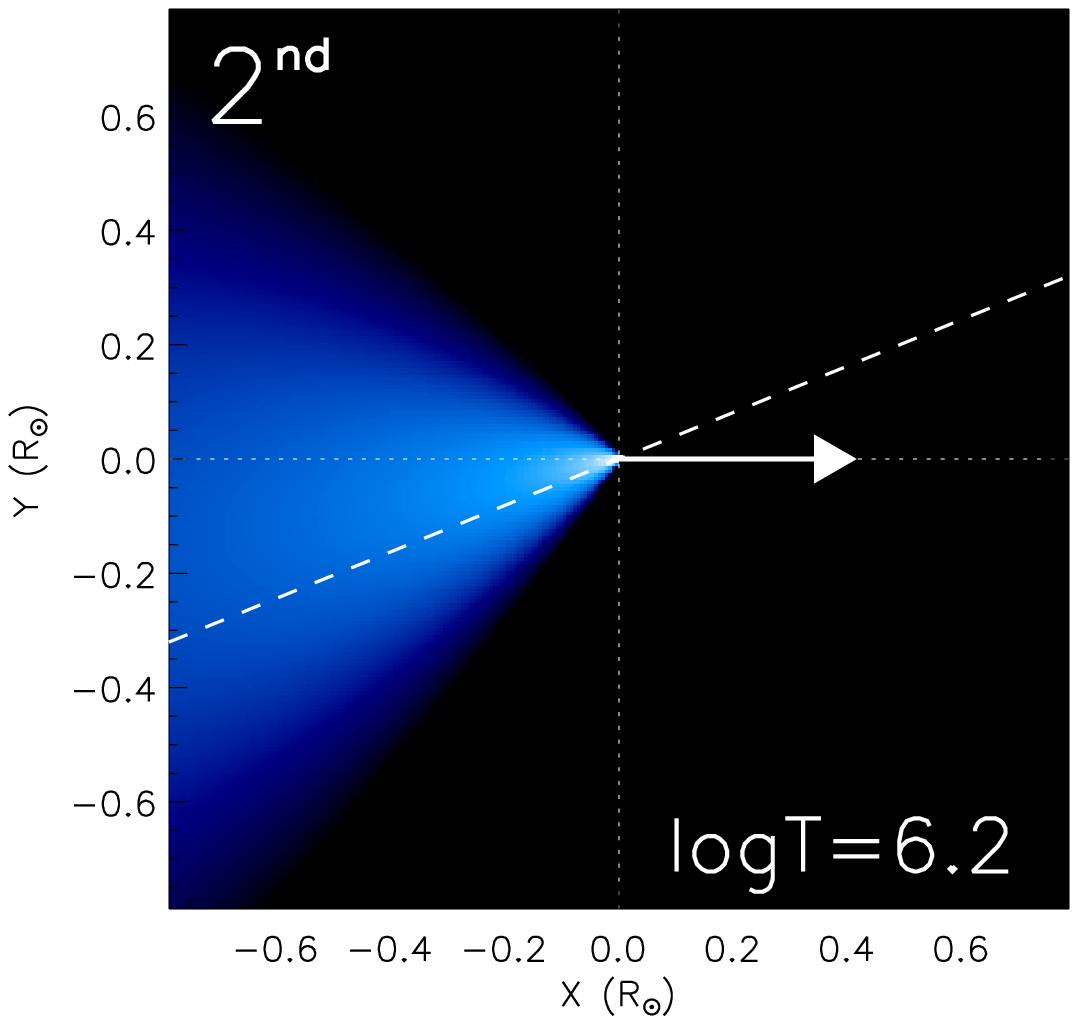}}  &
      \resizebox{60mm}{!}{\includegraphics{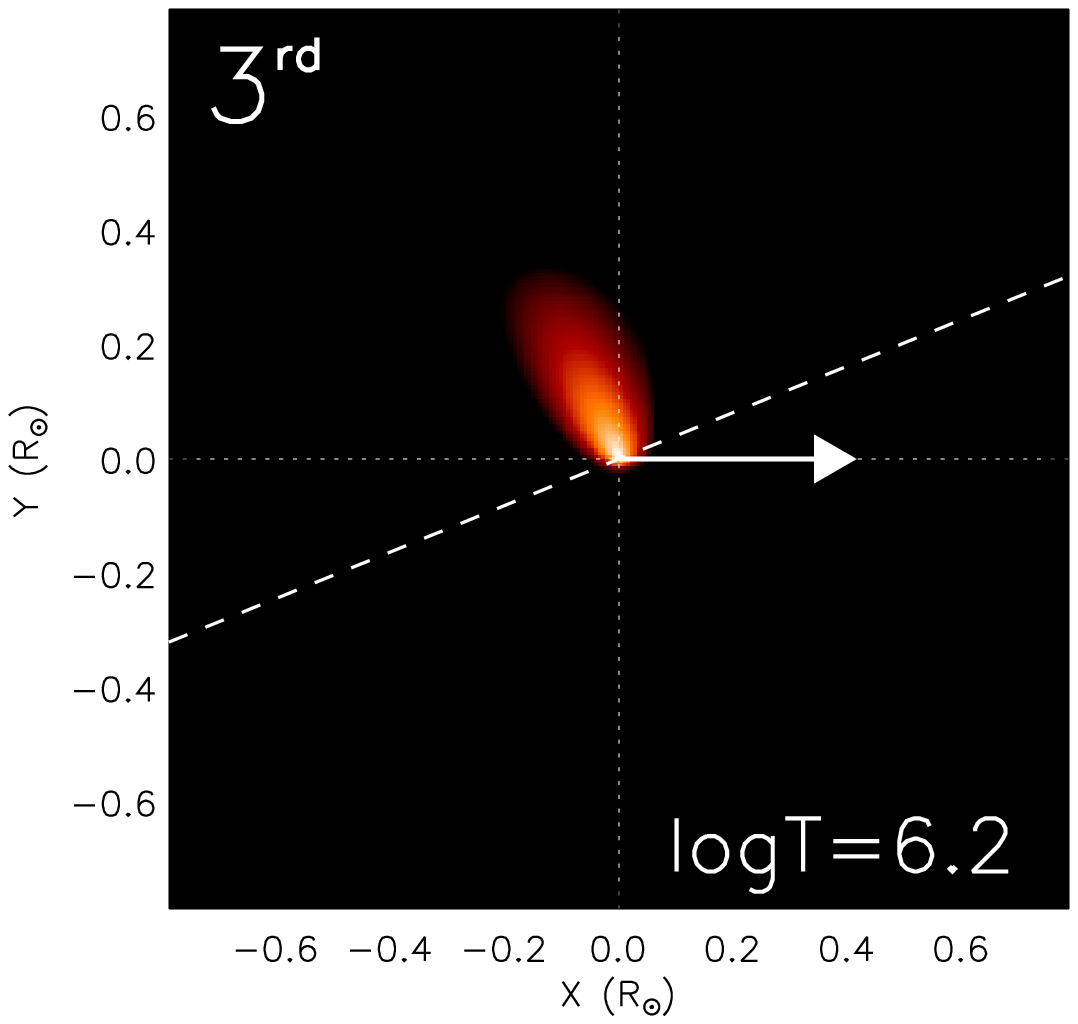}}  
    \end{tabular}
    \caption{Model of the neutral hydrogen densities integrated along the direction perpendicular to the comet trajectory from a comet at actual distance 8.00~\RSUN moving along the X-axis.
    The 1$^{st}$, 2$^{nd}$ and 3$^{rd}$ generation neutrals are shown in a frame of reference moving with the comet.  These model parameters are $\dot{N} = 1 \times 10^{29}~\rm s^{-1}$, a coronal density of $10^4~\rm cm^{-3}$, a wind speed of 200 \kms, and an angle between the comet trajectory and the magnetic field of 22.14$^{\circ}$, typical of Kreutz sungrazers in a radial magnetic field.  The scale of these images is shown in solar radii. The dashed lines show the radial direction and the arrows show the direction of the comet. The top panel show densities with coronal proton temperature $log T_{cor}=6.0$, the bottom $log T_{cor}=6.2$.
    \label{reference}}
  \end{center}
\end{figure*}

\noindent
Here $\dot{N}_{29}$ is the outgassing rate in units of $10^{29}$ H atoms per second, $n_4$ is the coronal density in units of $10^4~\rm cm^{-3}$, and $V_{10}$ is the outflow speed of the 1$^{st}$ generation neutrals in units of 10~\kms. These scalings break down at high outgassing rates when 3$^{rd}$ generation neutrals have a significant probability of experiencing another charge transfer and when the comet begins to significantly modify the structure of the corona. The former occurs at about $\dot{N}_{29}~\sim~3$, while the latter depends on both outgassing rate and coronal density.  There is also some dependence on $V_{rel}$, which enters the charge transfer rate, so the numbers above should be taken as estimates at the factor of two level.

Different viewing angles affect the observed \LA\/ emission in a straightforward way.  Because the comet's motion, the wind velocity and the bulk motion of the pickup ions all lie in the plane of the sky in the simulation, the Doppler velocities of all three components are zero when viewed side-on.  Figure~5 shows the total \LA\/ intensity and Doppler velocity, combining the emission from the 3 populations, integrated along the line of sight for the same model when viewed from different phase angles, $\alpha$ = 30$^\circ$, 60$^\circ$ and 90$^\circ$ from the side-on view.  
The main tail due to 2$^{nd}$ generation neutrals is foreshortened, as is the angle between the 3$^{rd}$ generation tail and the Z-axis.  The comet is moving away from the observer, giving a positive velocity for the 1$^{st}$ and 3$^{rd}$ generation emission, while the solar wind is coming toward the observer, giving a blue-shift.  The solar wind speed is taken to be 200~\kms.  The predictions for the other angles can be obtained by rotation, so that the image might be flipped left to right or red- and blue-shifts might be interchanged.
The phase angle of the comet C/2002 S2 is computed from the orbital parameters, then we rotate the model by and angle of 31.40$^\circ$ out of the plane of the sky before the integration along the line of sight to obtain the total intensity and Doppler velocity 2D maps, finally we rotate the maps by 34$^\circ$ clockwise to put the comet at the solar latitude as seen by UVCS. 
The maps obtained from the reference model of Figure~4 are shown in Figure~6.

The following figures illustrate the dependence of the \LA\/ emission on various parameters.  All of them use a rotation of 31.40$^\circ$ out of the plane of the sky, and all start from the same reference parameters, varying one parameter at a time.  

Figure~7 shows the effect of decreasing or increasing the outgassing rate by a factor of 3.  The higher $\dot{N}$ model shows a more prominent 3$^{rd}$ generation component, because of the number of 3$^{rd}$ generation neutrals scales as $\dot{N}^2$, as shown in equations 7 through 9.

Figure~8 shows models with wind speeds of 100 and 300~\kms.  The speed of the wind has 3 effects.  The faster wind makes Doppler dimming of the 2$^{nd}$ generation neutrals more severe, reducing the brightness of the 2$^{nd}$ generation component relative to the 1$^{st}$ and 3$^{rd}$ generations.  It also lengthens the 2$^{nd}$ generation tail, reducing the opening angle of the 2$^{nd}$ generation tail.  In addition, the axis of the 2$^{nd}$ generation tail falls between the comet trajectory and the wind direction, and it is closer to the comet trajectory at low wind speed and closer to the wind direction at high wind speed (Bemporad et al. 2015).

\section{Comparison with Observations}

\begin{figure*}[h]
  \begin{center}
    \begin{tabular}{cc}
      \resizebox{70mm}{!}{\includegraphics{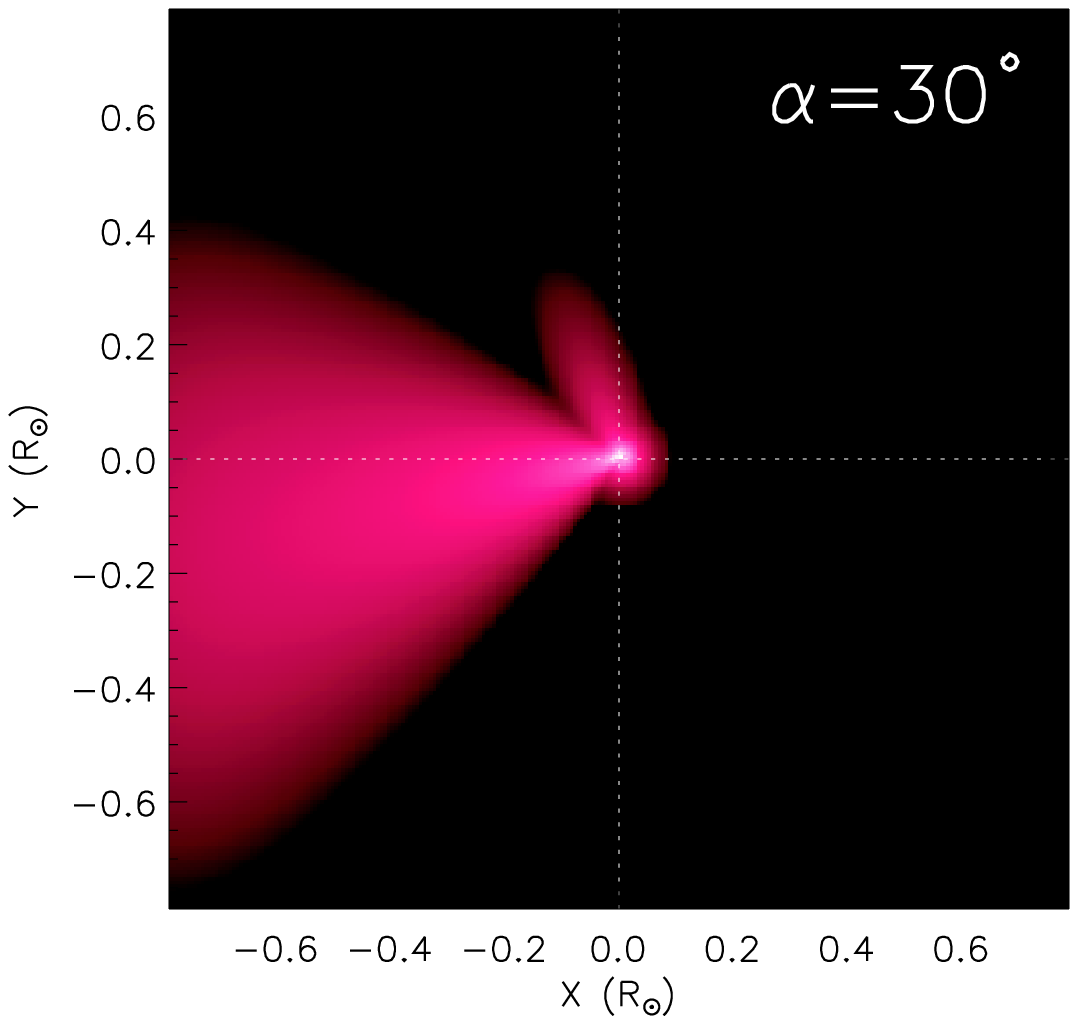}}  &
      \resizebox{70mm}{!}{\includegraphics{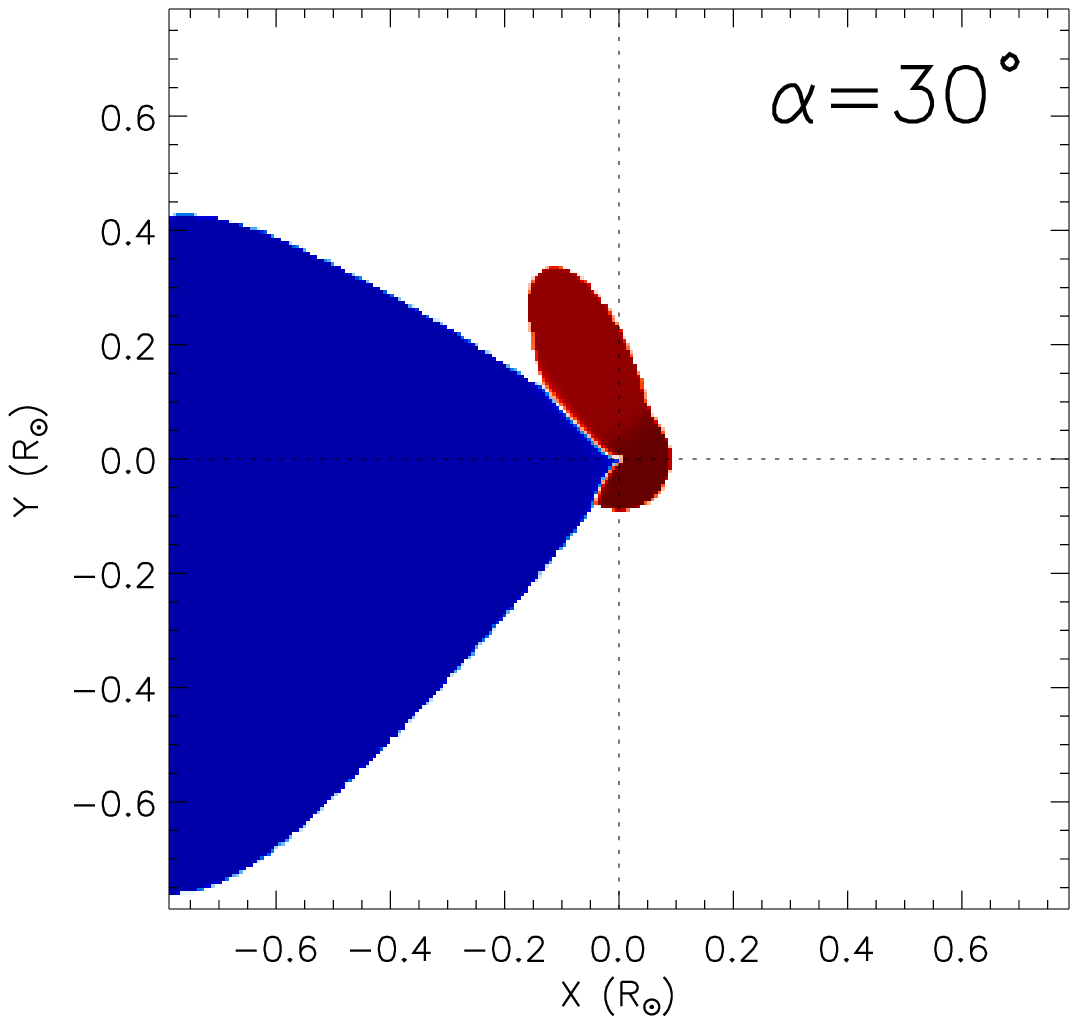}}  \\
      \resizebox{70mm}{!}{\includegraphics{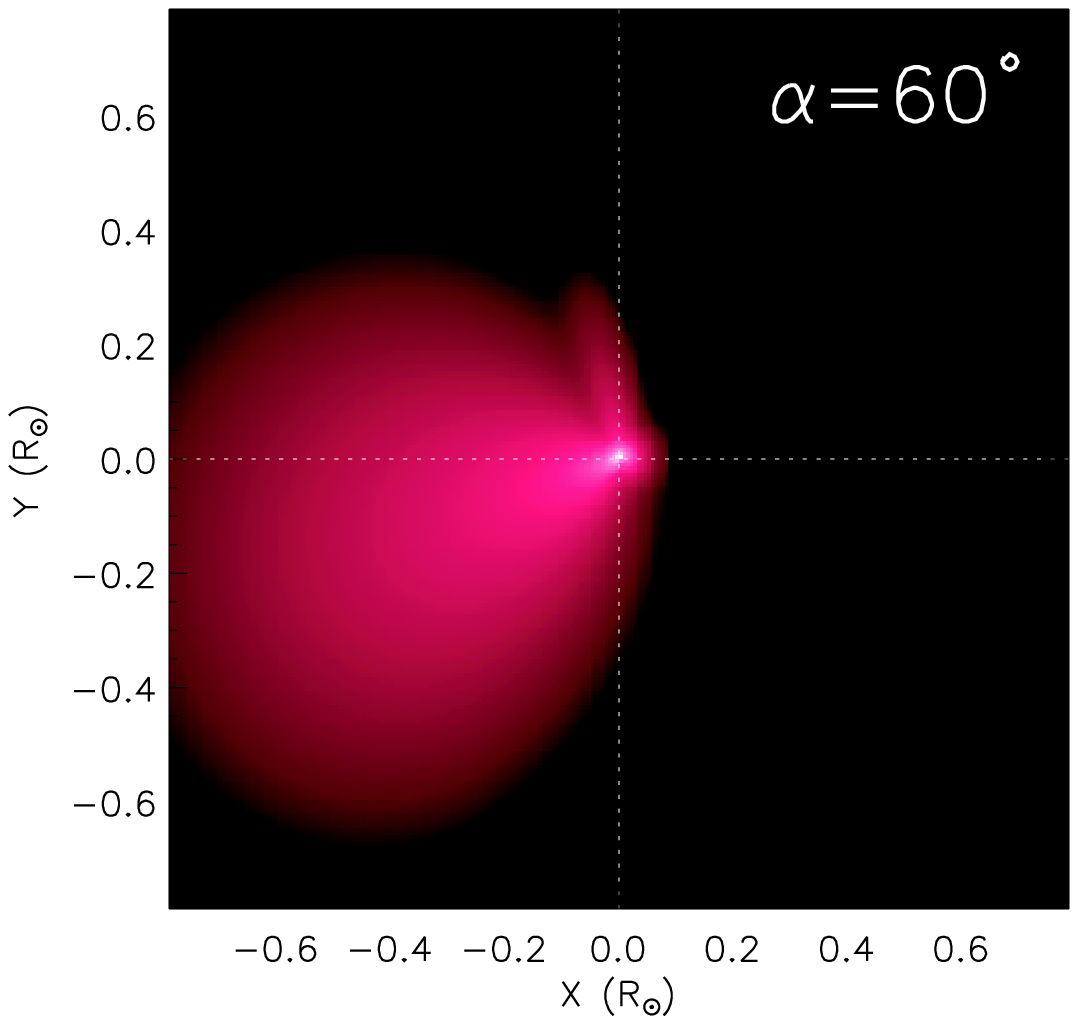}}  &
      \resizebox{70mm}{!}{\includegraphics{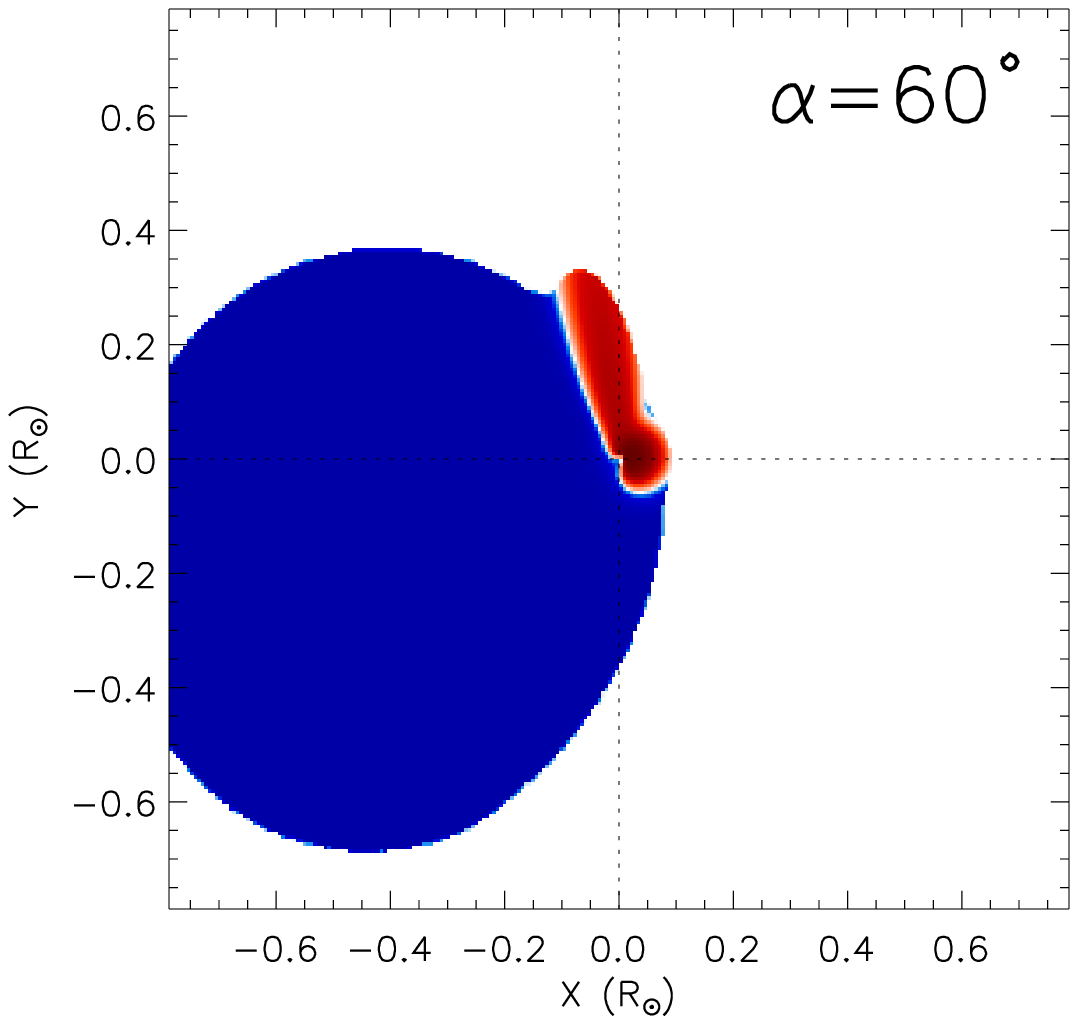}} \\
      \resizebox{70mm}{!}{\includegraphics{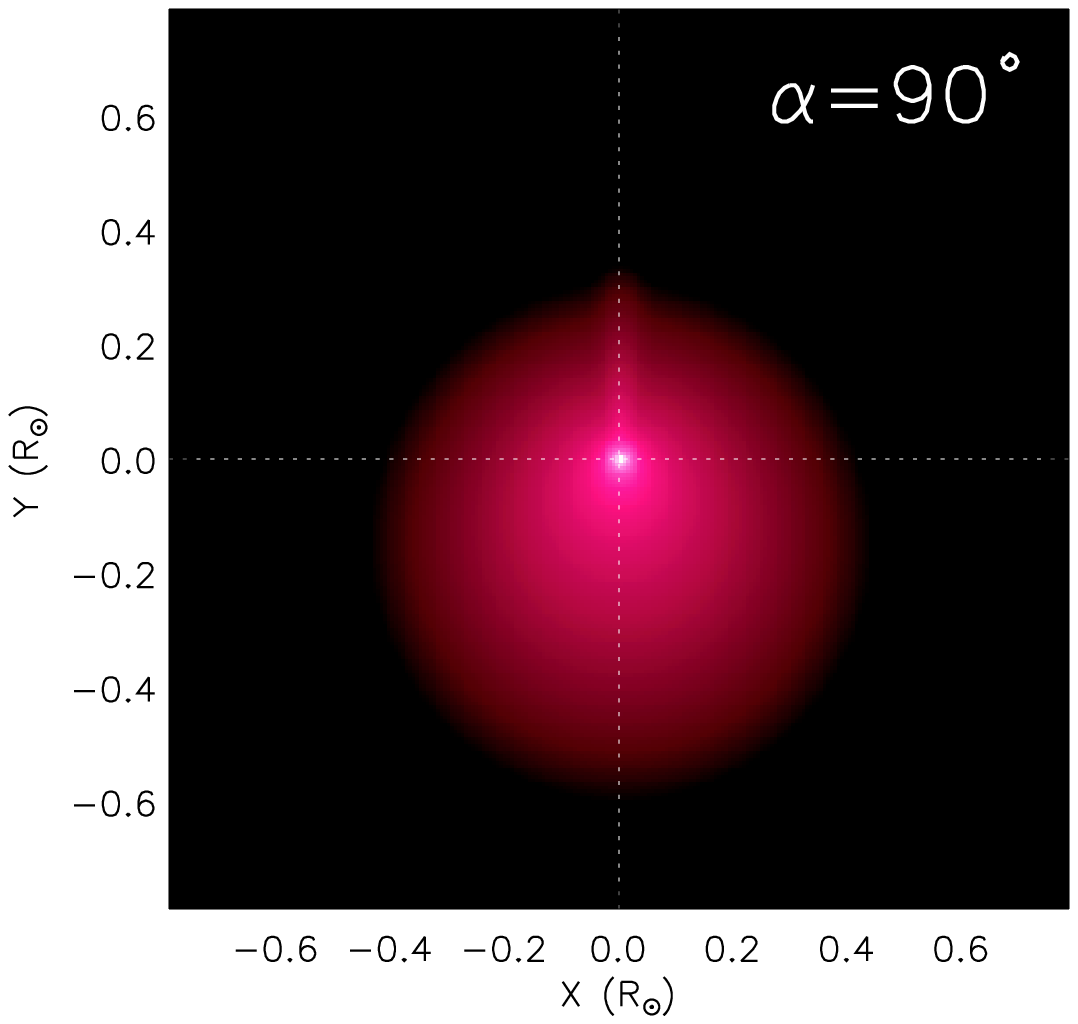}}  &
      \resizebox{70mm}{!}{\includegraphics{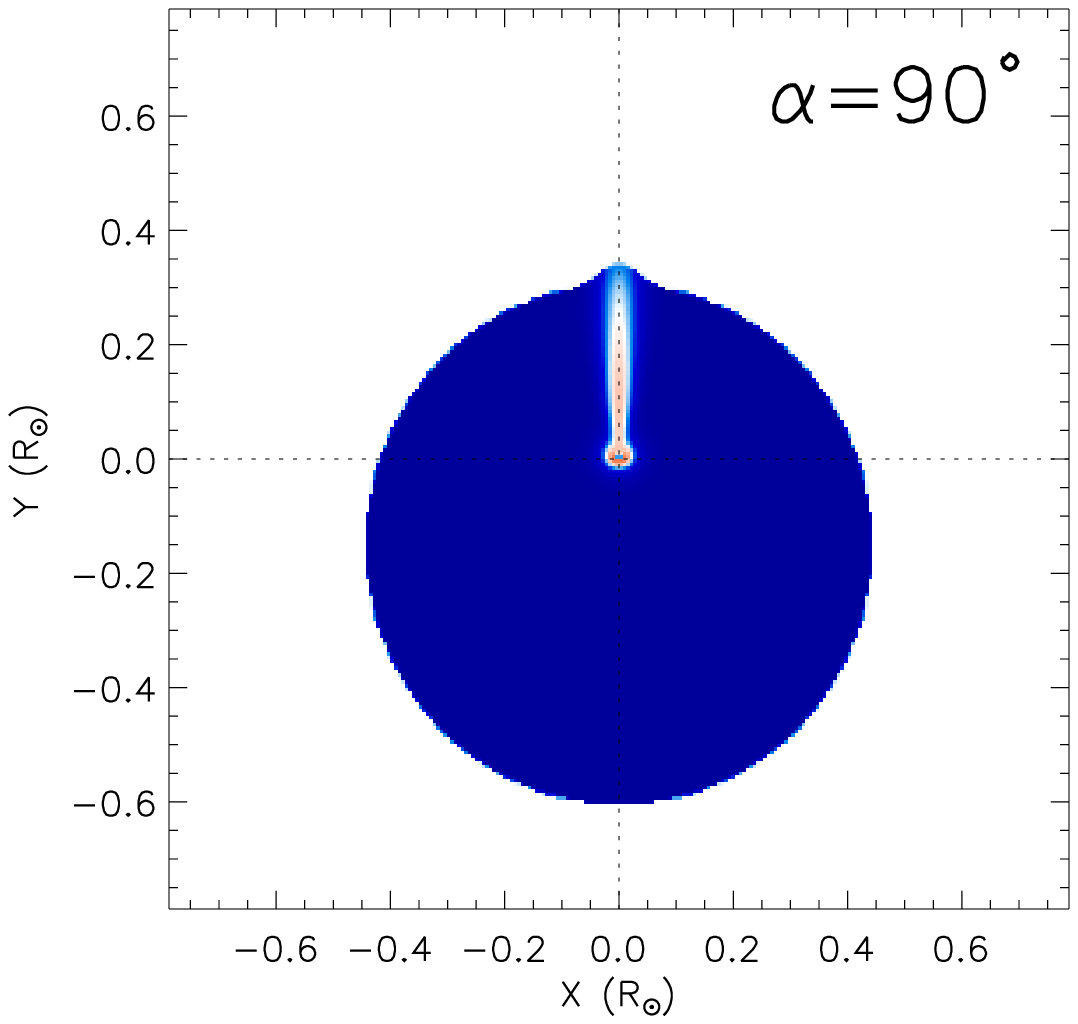}}  
    \end{tabular}
    \caption{Intensity and velocity images of the reference model, comet at actual distance 8.00~\RSUN and proton temperature log T=6.0, seen from angles of 30 degrees (top row), 60 degrees (middle row) and 90 degrees (bottom row).  Blue and red in the righthand panels correspond to blue-shifts and red-shifts as seen from Earth.
\label{rot_30_60_90}}
  \end{center}
\end{figure*}

Comet C/2002 S2 was analyzed by Giordano et al., (2015).  While the Ly$\alpha$ intensity maps reconstructed from the observation sequences were similar to the Monte Carlo models described in the paper, the intensity map for the 8~\RSUN crossing showed a split tail, with the southern side being much brighter (Fig.~1).  Worse, the observed Ly$\alpha$ centroids were red-shifted on one side of the tail and blue-shifted on the other (Fig.~2).  Giordano et al., (2015) suggested that the asymmetric appearance and the velocities could be explained if the blue-shifted emission arose from the 2$^{nd}$ generation neutrals and red-shifted emission from the 3$^{rd}$ generation.  In that case, the observed blue-shift is the LOS component of the solar wind speed and the red-shift is the LOS component of the parallel velocity of the PUIs, $V_\|$. 

We compare our simulation with the UVCS observations of Comet C/2000 S2 at 6 \RSUN and 8 \RSUN in Ly$\alpha$ (Giordano et al., 2015), attempting to match the brightness distribution and the Doppler shift asymmetry.  The observation at 7~\RSUN does not show as clear separation between red- and blue-shifted components.  Figure~9 displays the model for the 6 \RSUN crossing.  The model was computed as described above, then rotated to give its appearance projected onto the plane of the sky for the time of observation, and finally rotated 34 degrees to make the comet trajectory perpendicular to the UVCS slit, since that was the assumption used to reconstruct the images in Giordano et al., (2015). Finally, the predicted intensity is calculated by averaging brightness in a 300 arcsec box for comparison with the UVCS observations.

\subsection{Comparison at 6 \RSUN}

UVCS observed comet C/2002 S2 at 6 \RSUN from 23:17 until 00:16 UT on 18/19 September, 2002.  We assume the comet position and velocity vector from the orbit as given in Giordano et al. (2015), and we take the magnetic field to be radial.  That ought to be a good approximation between the top of the closed field arcades at the streamer cusps, which are up to about 4 \RSUN (Strachan et al 2002), and the Alfv\'{e}n radius at 10 to 20 \RSUN (Zhao \& Hoeksema, 2010, Deforest, Howard \& McComas, 2014 , Tasnim et al., 2018).  That assumption is critical because the angle between the comet trajectory and the field determines the perpendicular and parallel velocities of the PUIs and 3$^{rd}$ generation neutrals.  Moreover, the direction of the field determines the LOS components of the 2$^{nd}$ and 3$^{rd}$ generation neutrals. The solar wind velocity is not independently known, but to the extent that the comet orbit is reliable, we can take the agreement of the predicted and observed red-shifts of the 3$^{rd}$ generation emission as some support for the assumption that the field is radial.

\begin{figure*}
	\begin{center}
		\begin{tabular}{cc}
			\resizebox{70mm}{!}{\includegraphics{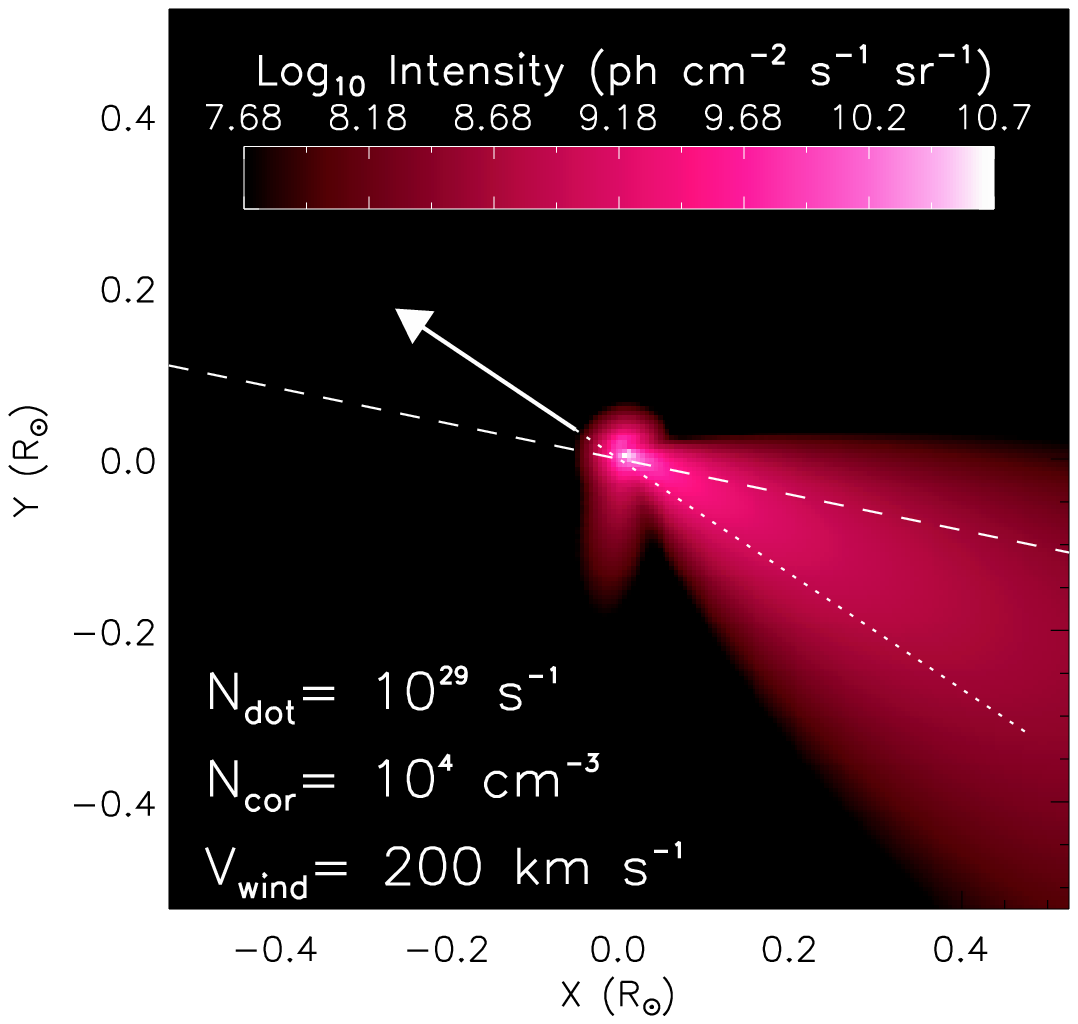}}  &  \resizebox{70mm}{!}{\includegraphics{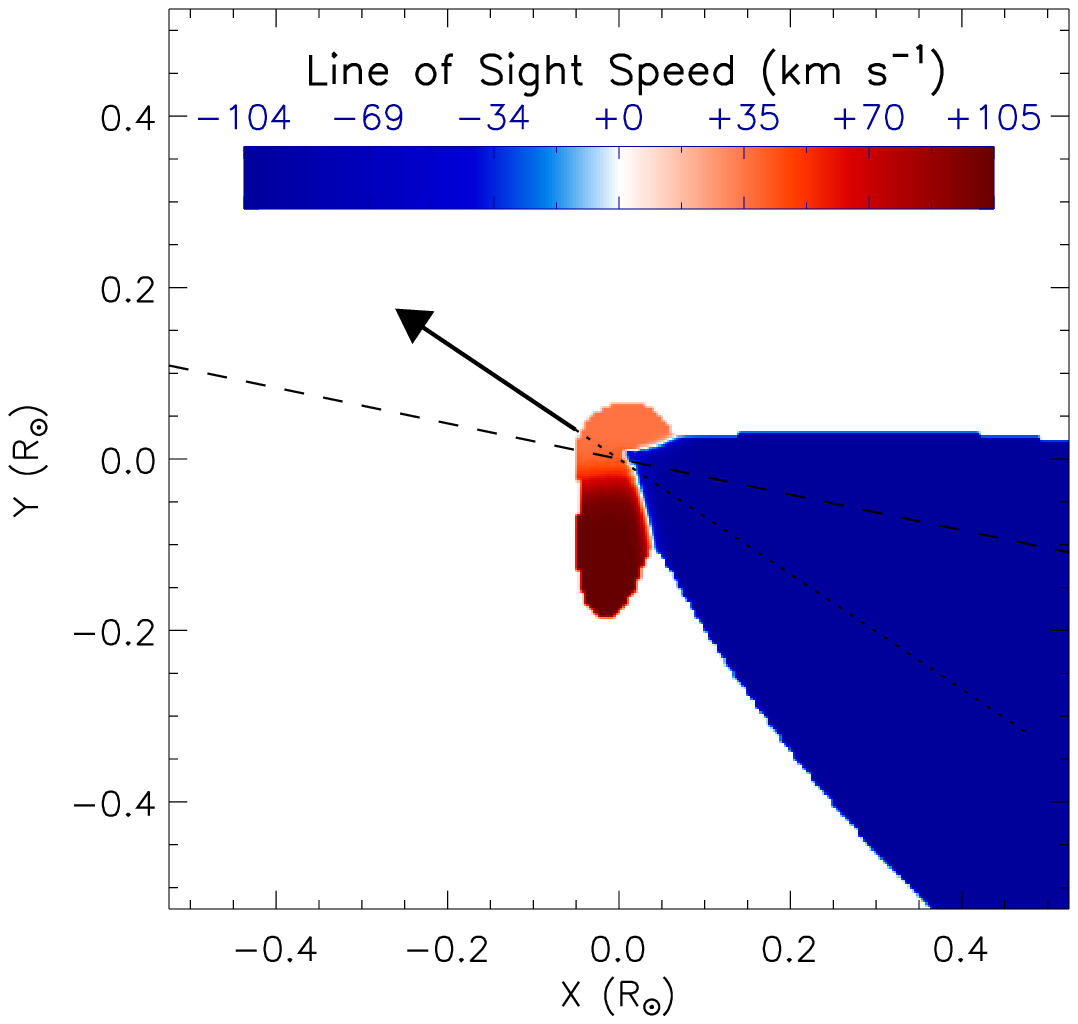}}  \\
		\end{tabular}
		\caption{Intensity and velocity images of the density model shown in Figure 4 seen from an angle of 31.40$^{\circ}$, 
		the comet is at actual distance 8.00~\RSUN and proton temperature log T=6.0. 
		The dashed lines show the radial direction and the arrows show the direction of the comet. 
		\label{rot_30}}
	\end{center}
\end{figure*}

\begin{figure*}
	\begin{center}
		\begin{tabular}{cc}
			\resizebox{70mm}{!}{\includegraphics{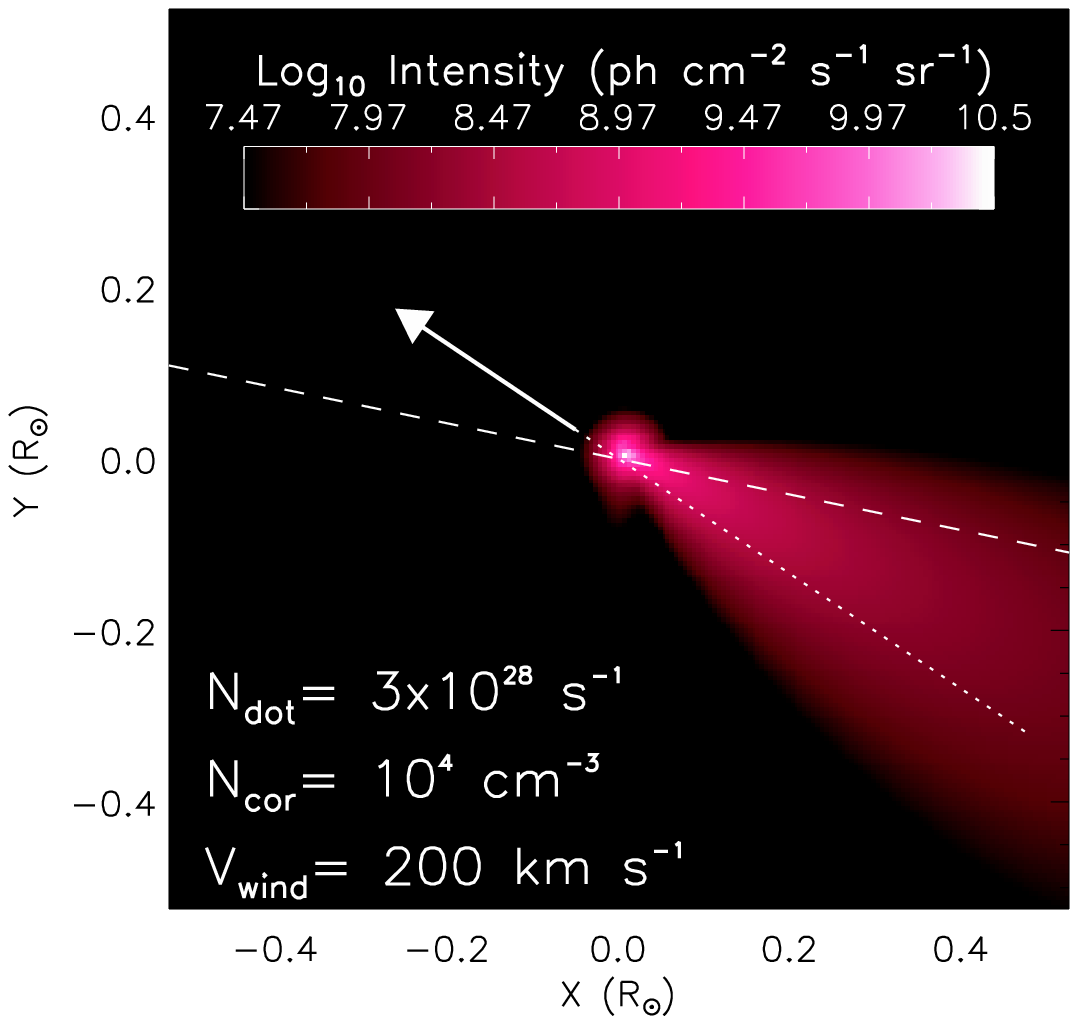}}  &  \resizebox{70mm}{!}{\includegraphics{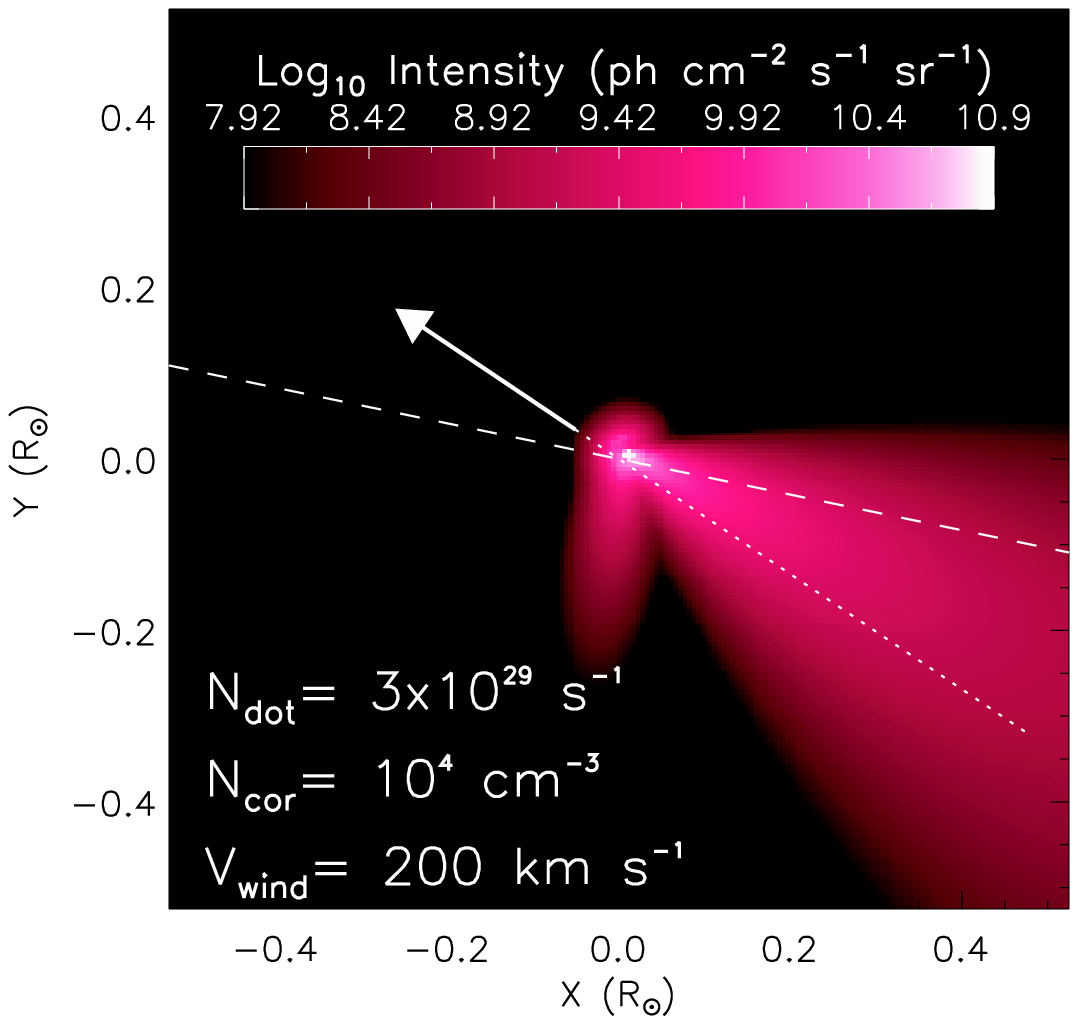}}  \\
		\end{tabular}
		\caption{Intensity images of models with outgassing rates 3 times smaller ($\dot{N} = 3 \times 10^{28}~\rm s^{-1}$, 
		left panel) and 3 times larger ($\dot{N} = 3 \times 10^{29}~\rm s^{-1}$, right panel) than the reference model.  
		This illustrates the increase in relative intensity of the 3$^{rd}$ generation component with increase $\dot{N}$. 
		The dashed lines show the radial direction and the arrows show the direction of the comet. 
		\label{Ndot-dependence}}
	\end{center}
\end{figure*}

We assume a solar wind velocity parallel to the magnetic field with a magnitude of 175 \kms.  This is much larger than the value of 75 \kms inferred from the shape of the  Ly$\alpha$ intensity image by Giordano et al. (2015), but we find that it is required to match the observed blue-shift of 111 to 145 \kms. This large velocity implies strong Doppler dimming, so a much larger mass loss rate, $1 \times 10^{29}$ H per second, is needed to match the Ly$\alpha$ brightness. We have no independent measurement of the wind velocity at this height, but Cho et al. (2018) give a range of about 130 to 430 km/s at 6 \RSUN \/ based on white light observations.  The density chiefly controls falloff of the 2$^{nd}$ generation intensity; the simulation with a coronal proton density of 10$^4 \rm cm^{-3}$, in agreement with Giordano et al. (2015), agrees reasonably well with the data (Figure~9). 

The assumed thermal velocity of {130}~\kms enters the Doppler dimming and the opening angle of the 2$^{nd}$ generation emission.  This corresponds to a kinetic temperature of log T={6.0}, however, the effective temperature including non-thermal (wave) motions could be larger.  Frazin, Cranmer \& Kohl (2003) report a 1/e width of Ly$\alpha$ in a streamer at 5.1~\RSUN (their largest height) of about 155~\kms.

Another poorly known parameter is the outflow speed from the comet.  We assume 10 \kms, though that is somewhat larger than is generally assumed.  The speed of the H atoms produced by photodissociation of H$_2$O ranges from about 8 to 24 \kms, but some fraction of that speed is lost to interactions with O atoms.  With smaller outflow speeds our simulation code does not resolve the region where 1$^{st}$ generation neutrals and PUIs coexist.  A smaller outflow speed would reduce the 1$^{st}$ generation intensity by increasing the optical depth of the neutral cloud formed by dissociation of water. It would also increase the number of 3$^{rd}$ generation neutrals because the PUIs would be formed in a region of higher neutral density.

Finally, there is some uncertainty in the orbit. That will mainly affect the projection of the cloud of Ly$\alpha$ onto the plane of the sky. In particular, a 10$^\circ$ to 15$^\circ$ error in the ascending node could significantly change the projected angle between the 2$^{nd}$ and 3$^{rd}$ generation clouds.

Comparison of Figure~6 with Figures~1 and 2 shows that the models get one essential aspect right.  The red-shifted 3$^{rd}$ generation emission lies to the south and the blue-shifted 2$^{nd}$ generation to the north, and their velocities agree with the observations.  However, the models do not really resemble the data.  The red-shifted 3$^{rd}$ generation emission appears as a short stub at a large angle to the main body of the comet tail in the model, while in the observations it is longer, and it makes a small enough angle to the main comet tail that it blends into the 2$^{nd}$ generation emission.  Because the predicted 3$^{rd}$ generation emission is so spatially confined, it creates the strong peak for about hundred seconds when the comet crosses the slit, but the sharp peak is not resolved in the observations because of the 120 second integration time of each exposure.

Figure~9 compares the Ly$\alpha$ intensity predicted by the model for 1$^{st}$, 2$^{nd}$ and 3${^rd}$ generation neutrals with the values observed by Giordano et al. (2015), and they match the observed intensities quite well.  The model parameters are different than those derived by Giordano et al. (2015), who assumed a slower wind speed.  The slower wind speed gives a good match to the opening angle of the Ly$\alpha$ tail, which then requires a higher density to match the falloff with time.  It now seems likely that the opening angle of the tail is determined by the separation between 2$^{nd}$ and 3$^{rd}$ generation tails, and the 2$^{nd}$ generation tail is narrower than we had thought, consistent with a faster solar wind.  The faster wind speed also implies more severe Doppler dimming of the 2$^{nd}$ generation emission, so our fit requires a higher outgassing rate of $10^{29}~\rm s^{-1}$, as opposed to $10^{28}~\rm s^{-1}$ from the earlier Monte Carlo simulation. The model in Figure~9 assumes a wind speed of 100 \kms, and a somewhat larger speed is needed to fully match the observed blue shift.

\begin{figure*}
	\begin{center}
		\begin{tabular}{cc}
			\resizebox{70mm}{!}{\includegraphics{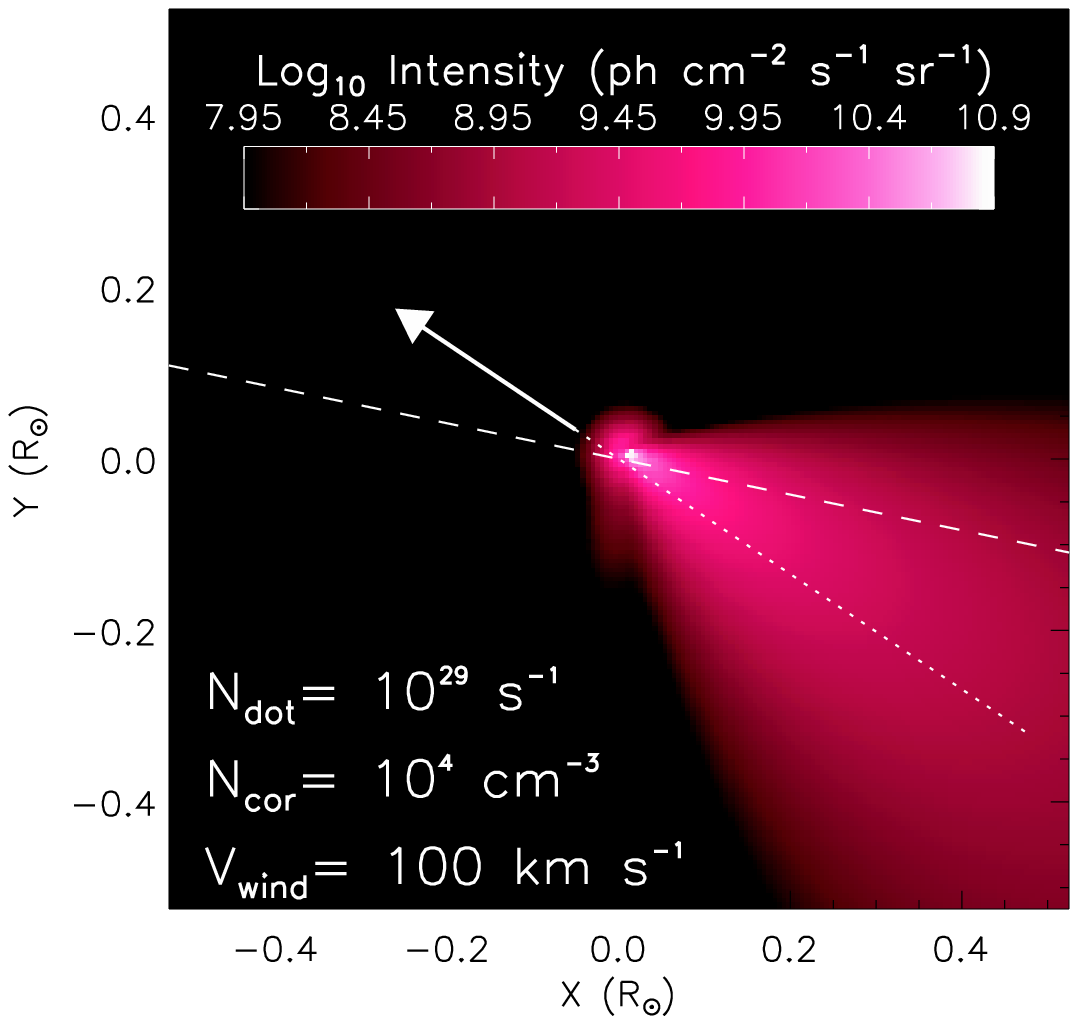}}  &  \resizebox{70mm}{!}{\includegraphics{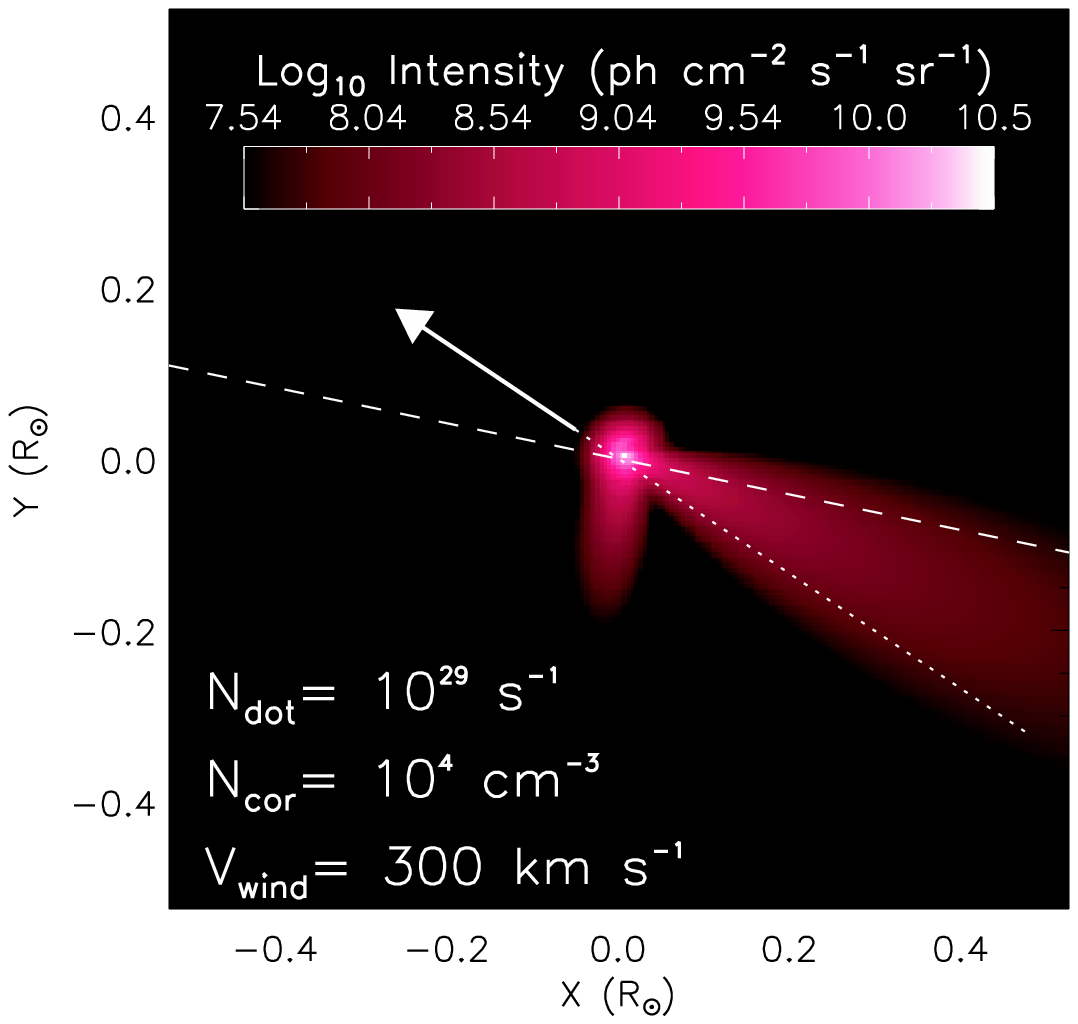}}  \\
		\end{tabular}
		\caption{Intensity images of the model with wind velocities of 100~\kms\, (left panel) and 300~\kms\, (right panel).  
		These models illustrate decreased intensities of the 2$^{nd}$ generation neutrals at high velocity due to Doppler dimming, 
		the longer and narrower 2$^{nd}$ generation component, and the closer alignment of the 2$^{nd}$ generation to the wind direction. 
		The dashed lines show the radial direction and the arrows show the direction of the comet. 
		\label{Vwind-dependence}}
	\end{center}
\end{figure*}

Overall, it appears that the basic idea of the 3$^{rd}$ generation emission is correct, in that it explains the otherwise baffling red- and blue-shifted emission from the south and north sides of the comet tail.  Part of the discrepancy may be that the observations missed some of the 1$^{st}$ and 3$^{rd}$-generation emission, because the comet was already in the UVCS slit during the first exposure.  On the other hand, there is clearly something wrong with either the model parameters or with the basic physics in the model.

The parameter most closely related to the appearance of the 3$^{rd}$ generation emission as a stub nearly at a right angle to the 2$^{nd}$ generation emission is the angle between the comet trajectory and the magnetic field.  When this angle is small, the PUIs move along the field at nearly the speed of the comet, which then means that they almost move with the comet and occupy a small interval relative to the comet before they are ionized.  As described above, the angle is set by the comet trajectory and the assumption that the field is radial.  The extremely elongated orbit and the small perihelion distance of the Kreutz sungrazers imply that the trajectory is not far from radial at 6 \RSUN,  and the angle is 25.7 degrees.  The magnetic field is expected to be close to radial in a steady model, but it could be perturbed by a CME. LASCO C2 difference images show a 450 \kms west limb CME at 08:30 UT and continuous low level activity in the SW quadrant, so it is possible that the field in 10-20$^\circ$ from radial.  We note that the LASCO images are dominated by plasma near the plane of the sky, while the comet was about 4 \RSUN away from the plane of the sky, so the LASCO movies may not accurately reflect conditions near the comet.

\subsection{Comparison at 8 \RSUN}

UVCS observed the comet at 8 \RSUN from 21:31 until 22:22 UT on 18 September, 2002.  The UVCS images at 8 \RSUN in Figures~1 and 2 show a split tail as well as red- and blue-shifts on the southern and northern sides of the tail.  The blue-shifted northern side is very faint, which is the expected result of Doppler dimming of the 2$^{nd}$ generation neutrals for a solar wind speed around 200 \kms. The split tail is also expected as a result of the angle between the 2$^{nd}$ and 3$^{rd}$ generation populations.  However, the observed brightness increases gradually over the course of about 15 minutes (Figure 5 of Giordano et al. 2015), while the models predict a bright, short-lived 3$^{rd}$ generation peak followed by a slowly fading 2$^{nd}$ generation tail.  The width of the tail during even the first UVCS exposure suggests that the observations may have missed the passage of the nucleus and 1$^{st}$ generation cloud across the slit, but that still would not explain the gradual brightening.

The models can easily match a 3$^{rd}$ generation tail brighter than the 2$^{nd}$ generation tail with a wind velocity high enough to give strong Doppler dimming.  However, the
models predict a large angle between the two tails and a 3$^{rd}$ generation tail that is bright for only a short time.
This is at least partly due to our approximate treatment of the 3$^{rd}$ generation neutrals as discussed below.

\subsection{Physical explanations for the discrepancies}

There are three apparent aspects of the physics in the model that might be problematic; the assumption that the pickup ions do not interact with the ambient medium, the assumption that we can ignore the effects of the comet on the ambient plasma and the assumption that we can treat each species as a single fluid when computing the Doppler dimming.

We have assumed that the pickup ions form with parallel and perpendicular velocities determined by the angle between the comet trajectory and the B field, and that $V_\|$ is conserved while $V_\bot$ is scattered into a shell centered on $V_\|$.  This assumption receives some support from the AIA observations of Comet Lovejoy (Raymond et al (2014)), though the AIA observation indicated that the shell in velocity space was quickly scattered into a Maxwellian.  However, for the modest angle between the orbit and the radial B field at 6 \RSUN, $V_\|$ is more than twice $V_\bot$, so that the PUIs may be subject to violent streaming instabilities.  Those instabilities can transfer momentum between the wind and the PUIs and they can heat both populations.  They can also create large amplitude fluctuations in the magnetic field, so that a range of $V_\|$ and $V_\bot$ is present.  Those effects will broaden the 3$^{rd}$ generation tail, push it closer to the 2$^{nd}$ generation tail and make it longer.  However, if the relative speed of the comet and the wind is less than the Alfv\'{e}n speed the turbulence may be weak, while if the interaction is too strong it will reduce the red-shift vanish, in contradiction to the observations.

We have also assumed that the mass flow from the comet does not greatly perturb the wind.  That assumption is likely to be good for the relatively small $\dot{N}$ found by Giordano et al. (2015), but it is somewhat questionable for the higher value indicated by our model.  The interaction can decelerate the wind, which would make it difficult to explain the strong blue-shift observed.  On the other hand, we do not independently know the wind speed, and we might be observing a flow that has been decelerated.  If the perturbation is significant, a bow shock can form in the mass-loaded wind (Gombosi et al. 1996, Jia et al. 2014), and the changes in velocity and temperature can be large.  Dramatic shifts of more than 100 \kms in the velocity centroid and line width can be seen at outgassing rates above $10^{30}~\rm s^{-1}$ in Comet Lovejoy (Raymond et al 2018).

The largest part of the discrepancy may be related to the perturbation of the wind by the comet. Behar et al. (2018a and 2018b) used analytic and numerical models to study the interaction between the outflowing gas of comet 67P/Churyumov-Gerasimenko and the solar wind.  In that case, the Larmor radii of the ions are comparable to the size of the interaction region, and significant electric fields arise.  In the case of C2/2002 S2, even the Larmor radius of the O$^+$ ions is only a percent or so of the size of the interaction region, but the formation of 3$^{rd}$ generation neutrals is strongly biased toward small distances from the comet because of the rapid density falloff. Behar et al. (2018b)  showed that cometary ions are deflected in one direction and solar wind ions in the other, which would tend to produce the split tail we observe and separate the 2$^{nd}$ and 3$^{rd}$ generation neutrals.  In the case of comet C/2002 S2, the main effect may be that the comet deflects the magnetic field.  That both increases the angle between the field and the wind and deflects the wind away from its radial flow.  The former increases the perpendicular component of the pickup ion velocity, reduces the parallel component, and produces a more elongated tail in the direction of the comet motion, producing a red-shifted component more like that observed.  The latter changes the angle between the solar wind and the LOS, decreasing the wind speed needed to match the observed LOS component.  That helps with severe Doppler dimming expected from the observed blue-shift if an unperturbed radial flow is assumed, and it therefore, reduces the required outgassing rate.

Finally, we have computed the Doppler dimming under the assumption that each component can be characterized by a Maxwellian velocity distribution or a shell in velocity space with a given width and centroid.  This is a reasonable approach if there is no correlation between velocity and position, but it can break down for a non-interacting cloud of expanding neutrals.  In particular, the 3$^{rd}$ generation neutrals have a substantial velocity component toward the Sun and a somewhat narrower velocity width.  Atoms that have relatively small velocities toward the Sun will stretch farther along the tail of the comet, and they will also experience relatively weak Doppler dimming.  This coupling of the velocity and Ly$\alpha$ emission may explain the gradual intensity rise in the 8 \RSUN observation, since the H atoms farther behind the comet scatter Ly$\alpha$ photons from the disk very effectively, while H atoms that remain close to the comet experience stronger Doppler dimming.  

\begin{figure*}[h!]
	\centering
	\includegraphics[width=0.50\textwidth,angle=90]{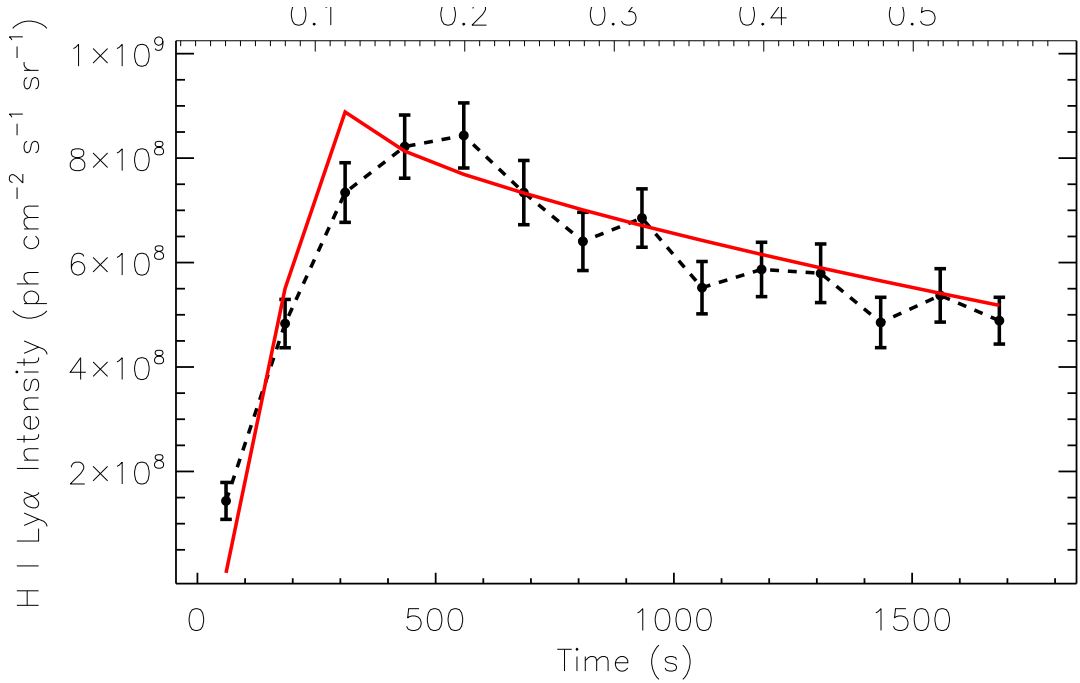}
	\caption{Comparison of the Ly$\alpha$ light curve of Comet C/2002 S2 at actual distance 6.00~\RSUN from Giordano et al., (2015) (black dots with error bars) 
	with predicted emission for the best fitting 3 generation model determined with an outgassing rate $\dot{N} = 1~\times~10^{29}$ H per second, 
	coronal density $n_{p}=1~\times~10^{4}~\rm~cm^{-3}$ and wind speed $V_{wind}=$100~\kms. 
	There is some uncertainty in the outgassing rate because of undertainty in the photoionization rate due to solar activity.}
	\label{S2_model_hist}
\end{figure*}

\section{Summary}

A 3$^{rd}$ generation of neutral H atoms formed by charge transfer of PUIs with cometary neutrals seems to basically explain the red-shifted Ly$\alpha$ emission from one side of the tail of comet C/2002 S2, while the blue-shift of the northern part of the Ly$\alpha$ can be understood as the LOS component of the 2$^{nd}$ generation neutrals moving at the solar wind speed. The existence of both 2$^{nd}$ and 3$^{rd}$ generation neutrals moving in different directions along the magnetic field lines can also explain the split tail observed at 8 \RSUN.

Our models do not match the observations in detail in that they predict too large an angle between the 2$^{nd}$ and 3$^{rd}$ generation tails, too short a 2$^{nd}$ generation tail, and no gradual increase in the 2$^{nd}$ generation brightness.  This may be partly due to incorrect parameter choices, such as the assumption that the magnetic field is radial, but it is probably also due to our approximate treatment of each generation of neutrals as a single velocity distribution independent of position.  To remedy the latter problem, we will need a kinetic simulation like that of Giordano et al. (2015) that includes the 3$^{rd}$ generation.

In any case, it is clear that the total emission of comet C/2002 S2 is dominated by the 3$^{rd}$ generation neutrals at 8~\RSUN, so the models used for other comets would not apply and parameters derived from those models should not be trusted.  On the other hand, the emission from that comet at 6~\RSUN is dominated by the 2$^{nd}$ generation. Parameters derived from the earlier models should in principle be valid, though in this case, the earlier model assumed a wind speed too small to explain the observed Doppler velocity. 

We note that an asymmetric, split tail in comet C/2001 C2 at 4.98~\RSUN was interpreted by Bemporad et al., (2005) in terms of fragments of a larger body that had not yet separated very far. That observation may well be amenable to interpretation in terms of 2$^{nd}$ and 3$^{rd}$ generation neutrals. Figure 6 of Bemporad et al., (2005)  shows evidence for changes in the Ly$\alpha$ line profile with time, which would support that interpretation.   

The Metis instrument on the Solar Orbiter spacecraft will provide Ly$\alpha$ images of comets, and Bemporad et al., (2015) have discussed how those images can be used to derive the density, temperature an outflow speed of the solar wind.  The presence of 3$^{rd}$ generation neutrals both complicates the interpretation and presents the opportunity to extract further information from the images.  In advantageous cases where the 2$^{nd}$ and 3$^{rd}$ generation are clearly resolved,  it will be possible to learn more about the magnetic field direction and the outflow speed.  In unfavorable cases where the components are projected on top of each other, it may make the interpretation more ambiguous. 

Overall, our models prove the concept that a 3$^{rd}$ generation of neutrals produced from PUIs can explain the morphology and velocity structure of comet C/2002 S2, but more sophisticated models are required to match the observations in detail and to extract more reliable physical parameters for the comet and the solar wind.  In particular, while it may be possible to include the perturbation of the wind and magnetic field to some extent in simple models like those presented here, a model that includes MHD and perhaps ion kinetic effects is probably needed.

\acknowledgments
This work was supported by the NASA LWS program under ROSES
NNH13ZDA001N. It benefited greatly from the workshop on near-Sun comets at the International Space Science Institute in Bern, Switzerland led by G. Jones. 

\facilities{SOHO (UVCS) }

\clearpage

\end{document}